\documentclass{emulateapj}

\newcommand{\icarus}{Icarus}

\begin{document}


\title{The Demographics of Long-Period Comets}


\author{Paul J. Francis}
\affil{Research School of Astronomy and Astrophysics, the Australian
National University, Mt STromlo Observatory, Cotter Road, Weston ACT 2611, Australia}
\email{pfrancis@mso.anu.edu.au}

\begin{abstract}
The absolute magnitude and perihelion distributions of long-period comets are
derived, using data from the Lincoln Near-Earth Asteroid Research (LINEAR) survey. 
The results are surprising in three ways. Firstly, the flux of comets through the inner
solar system is much lower than some previous estimates. Secondly, the expected rise in
comet numbers to larger perihelia is not seen. Thirdly, the number of comets per unit
absolute magnitude does not significantly rise to fainter magnitudes. These results imply that
the Oort cloud contains many fewer comets than some previous estimates, that small long-period
comets collide with the Earth too infrequently to be a plausible source of Tunguska-style 
impacts, and that some physical process must have prevented small icy planetesmals from reaching 
the Oort cloud, or have rendered them unobservable. A tight limit is placed on the
space density of interstellar comets, but the predicted space density is lower still.
The number of long-period comets that will be discovered by telescopes such as SkyMapper,
Pan-Starrs and LSST is predicted, and the optimum observing strategy discussed.
\end{abstract}


\section{Introduction}

The Oort cloud \citep{oort50} remains the most mysterious part of our solar system, primarily
because it cannot be directly observed. Our only observational clues to the size, shape, mass 
and composition of the Oort cloud come from observations of long-period comets. The
demographics of observed long-period comets have been the starting point of almost all
attempts to model the Oort cloud \citep[eg.][]{oort50,wei96,wei99,don04}.

Until the last ten or so years, the vast majority of comets were discovered by systematic
eyeball searches, using small telescopes \citep{hug01}. These surveys have been highly effective
at identifying large samples of comets, and in deriving their orbital parameters. They do, however,
have three major drawbacks:

\begin{itemize}

\item Unknown selection function: it is very unclear how often different parts of the sky
are surveyed, and to what depth. Surveys are clearly more sensitive to comets with bright
absolute magnitudes and perihelia close to the Earth, but the strength of this effect is
very hard to estimate \citep{eve67a,eve67}.

\item Limited range of comets observed: eyeball surveys find few comets fainter than an
absolute magnitude of 10 and with perihelia beyond 3AU.

\item Poorly defined photometry: these surveys quote the ``total brightness'' of a comet.
Total cometary magnitudes are notoriously unreliable. They are typically measured by defocussing 
a standard
star to the same apparent size as the comet, but this apparent size is heavily dependent on
observing conditions and observational set-up.

\end{itemize}

Despite these drawbacks, many attempts have been made to derive the basic parameters of
the long period comet population from eyeball-selected historic samples. The most heroic
and influential attempt was that of \citet{eve67}. Everhart carried out an exhaustive
analysis of the historical circumstances in which comets were discovered, over a 127 year
period. He developed a model for the sensitivity of the human eye, and used it to calculate
the period over which a given historical comet could have been seen. This was then used to
estimate the completeness of the comet sample: if a given type of comet was typically seen
early in its visibility window, surveys should be complete for this type of comet.
If the mean time to find a given type of comet is, however, comparable to the 
length of the estimated visibility window, the completeness is probably low. Using this method,
Everhart estimated that for every comet seen, another 31 were missed.

A more modest and recent attempt was that of \citet{hug01}. He restricted himself to the brightest
and nearest comets, for which he claimed (on the basis of discovery trends)
historical surveys were highly complete. The statistics of these comets were simply extrapolated
to larger perihelia and fainter absolute magnitudes, with no correction for observational 
incompleteness. As one would expect, the flux of long-period comets through the inner solar 
system estimated by \citet{hug01} is much lower that that estimated by \citet{eve67}.

Despite these attempts, several basic questions about the demographics of long-period
comets remain unresolved. One question concerns small comets \citep{bra96}: those with nuclear radii
less than $\sim 1$km (absolute magnitudes $H \ga 10$). Extrapolating the Everhart data
implies that there should be a large population of such comets. \citet{hug01} was
unable to tell whether his model predicted a large population of such comets or not. A second
question concerns the number of comets per unit perihelion. \citet{eve67} found that this
number rises from the sun out to 1AU, but was unable to determine whether it keeps rising at
larger perihelia. \cite{hug01} found no significant rise, but had large enough error bars to
bracket both of Everhart's possibilities.

The observational situation has changed radically in the last few years. The advent
of large format sensitive CCDs has allowed automated surveys to supplant eyeball searches as
the main mechanism for finding new long-period 
comets\footnote{\url{http://comethunter.de/}}. 
Most long period comets are now being
found as by-products of various automated searches for near-Earth objects, such as the 
Lincoln Near-Earth Asteroid
Research (LINEAR) project \citep{sto00}, the 
Catalina Sky Survey\footnote{\url{http://www.lpl.arizona.edu/css/}}, 
LONEOS\footnote{\url{http://asteroid.lowell.edu/asteroid/loneos/loneos.html}}
and NEAT\footnote{\url{http://neat.jpl.nasa.gov/}} \citep{pra99}. Many are also found by space-based
coronagraphs as they approach very close to the Sun \citep{bie02}, though these are mostly 
fragments of recently disintegrated larger comets \citep{sek04}.

In this paper, I attempt to deduce the statistical properties of the long-period comet 
population from one of these CCD surveys: the LINEAR survey. This has a far better defined
selection criterion than any historical eyeball survey, and extends to much larger perihelia
and fainter absolute magnitudes. It thus allows both an independent check and an extension of
previous estimates of the long-period comet population.

Near Earth asteroid (NEO) surveys are not optimized for comet detection. While they find
many long-period comets, they do not publish their raw data, nor all the details one would
like of their exact detection algorithms and sky coverage. In particular, they do not
publish on-going photometry of the comets they discover. Nonetheless, enough information is
available to make a first pass at estimating the true population of long-period comets from
their data. There have been previous attempts to use these surveys to detemine the true
populations of NEOs \citep[eg.][]{jed03} and dormant comets \citep{lev02}, but this paper is the
first attempt of which I am aware to do this for active comets.

In the next few years, the situation should further improve, with the advent of a new
generation of wide-field survey telescopes, such as 
SkyMapper\footnote{\url{http://www.mso.anu.edu.au/skymapper/}}, Pan-Starrs \citep{hod04} and 
Gaia 
\citep{per01}.
These surveys will predominantly find comets much fainter and more distant than historical
surveys. The analysis in this paper allows a first estimate of just how many long-period comets
these surveys can find, and how best to identify them.

I start off by defining a sample of comets drawn from the LINEAR sample,
and examining its properties, which are very different from those of eyeball samples (\S~\ref{sample}).
A model of the long-period comet population is then generated (\S~\ref{model}) based on and extrapolating 
the historical eyeball-selected surveys. A Monte-Carlo simulation of this comet population as 
it would be observed by LINEAR is then developed (\S~\ref{montecarlo}). The results are
compared to the observed sample in \S~\ref{compmodels}: I find that the Hughes model is
quite a good fit to the data, but that the Everhart model is not. I derive my own best-fit
model of long-period comet demographics. The consequences of this new model are many: I examine them
in \S~\ref{discussion} before drawing conclusions in \S~\ref{conc}.

\section{The Comet Sample}

\label{sample}

Of the several near-Earth asteroid surveys now under-way, the Lincoln Near-Earth Asteroid
Research (LINEAR) project \citep{sto00} was most suitable for constraining the
long-period comet population. This is because:

\begin{itemize}
\item They discover more comets than any other single survey.
\item They publish sky charts on their web page\footnote{\url{http://www.ll.mit.edu/LINEAR/}} showing
the area of the sky observed during each lunation, with the point-source magnitude limit
reached at each location.
\item Their sky coverage and magnitude limit is relatively simple and uniform across this period.
\end{itemize}

The comet sample was defined as follows:

\begin{enumerate}

\item The comet has an orbital period longer than 200 years.

\item The comet reached perihelion between 2000 Jan 1 and 2002 Dec 31.

\item The comet was either discovered by LINEAR between these dates, or could have been
discovered by LINEAR between these dates had it not already been discovered by someone else, 
or discovered prior to 2000 Jan 1.

\end{enumerate}

The 2000-2002 date range was chosen because comet details \citep[from the Catalog of 
Cometary Orbits,][]{mar03} and sky-maps (including limiting magnitudes) are available.

\citet{mar03} listed 25 comets as having been discovered or co-discovered by LINEAR which 
met our criteria. I needed, however, to add two additional sub-samples:

\begin{itemize}

\item Comets discovered prior to 2000, but which reach perihelion in the period 2000-2002, and 
which could have been first discovered by LINEAR within this period, had they not already
been found.

\item Comets discovered in 2000-2002 inclusive by other surveys, but which would subsequently
have been seen by LINEAR during this period.

\end{itemize}

Potential members of the two additional sub-samples were selected from \citet{mar03}. Each
candidate was checked for its detectability by LINEAR, using the ephemerides and predicted 
magnitudes generated by the Minor
Planet Center\footnote{\url{http://cfa-www.harvard.edu/iau/mpc.html}}.  
The predicted positions and
brightnesses were compared to the maps of LINEAR sky coverage. These maps show only the
integrated coverage per lunation, not the night-by-night or hour-by-hour coverage, but
most of these comets move slowly enough that this shouldn't much matter.

This process added another 27 comets to our sample. 8 had been detected by LINEAR during 1999,
but reached perihelion in 2000 or 2001. Most of the remainder were first identified by other 
near-Earth asteroid surveys, particularly the 
Catalina Sky Survey, LONEOS and NEAT. 

For every comet in our final sample, the original discovery details (as distributed by the
Central Bureau of Astronomical Telegrams) were checked. From these, the discovery date, discovery
magnitude $H_{\rm dis}$ and discovery circumstances were noted. The discovery magnitudes are total 
magnitudes (m1). It is not clear how reliable and homogeneous these magnitudes are, but no better
source of CCD photometry is available. They are based on CCD observations by professional 
astronomers of typically barely resolved objects, and so should be good to $\sim 0.5$mag.

Absolute magnitudes $H$ were computed from these discovery magnitudes $H_{\rm dis}$. The standard 
equation
was used:
\begin{equation}
H_{\rm dis} = H + 5 \log_{10} \Delta + 2.5 n \log_{10} r
\label{eqabsmag}
\end{equation} 
\citep[eg.][]{whi78}, 
where $H_{\rm dis}$ is the observed total magnitude at discovery and $n$ a power-law 
parameterization of 
the dependence on heliocentric distance. As is conventional for solar system work, the absolute
magnitude is defined as the observed magnitude if the object were at a distance of 1 AU from
both the Earth and the Sun.
Following \citet{whi78}, the dynamically new and old comets were treated 
differently (the new ones are much brighter at large heliocentric radii, at least on their way in).
A comet is classed as dynamically new if its original semi-major 
axis $a$ is $>$ 10,000 AU, old if $a <$ 10,000 AU, and undetermined if
the orbit class in \citet{mar03} is II or worse. For new comets, $n=2.44$ was used if they are seen
pre-perihelion and $n=3.35$ if seen afterward. For old comets, the values are 5.0 and 3.5
respectively. The canonical value of $n=4$ is used for comets of undetermined orbit type.
This is uncertain both because real comets show a dispersion in $n$, and
because the $n$ values in \citet{whi78} are based on observations at smaller
heliocentric distances. It is, however, self-consistent with the analysis used in our Monte-Carlo
simulations.

Our sample is listed in Table~\ref{sampletab}.

\begin{deluxetable}{lcccc}
\tablewidth{0pt}
\tablecolumns{5}
\tablecaption{The LINEAR Long-Period Comet Sample \label{sampletab}}
\tablehead{
\colhead{Name} & 
\colhead{q (AU)} &
\colhead{H} &
\colhead{1/a\tablenotemark{a}} &
\colhead{Orbit class\tablenotemark{b}}  
\\
}
\startdata

  C/1999 F1  &   5.7869  &  7.82 &  0.000038 & 1A \\
  C/1999 J2  &   7.1098  &  6.39 &  0.000019 & 1A \\
  C/1999 K5  &   3.2558  &  9.75 &  0.000024 & 1A \\
  C/1999 K8  &   4.2005  &  6.33 &  0.000681 & 1A \\
  C/1999 L3  &   1.9889  & 10.21 &  0.013741 & 1B \\
  C/1999 N4  &   5.5047  &  9.99 &  0.000068 & 1A \\
  C/1999 S4  &   0.7651  &  7.84 &  0.000720 & II \\
  C/1999 T1  &   1.1717  &  4.36 &  0.000173 & II \\
  C/1999 T2  &   3.0374  &  6.05 &  0.000596 & 1A \\
  C/1999 T3  &   5.3657  &  5.12 &  0.000231 & 1B \\
  C/1999 U4  &   4.9153  &  7.60 &  0.000037 & 1A \\
  C/1999 Y1  &   3.0912  &  9.80 &  0.000044 & 1A \\
  C/2000 A1  &   9.7431  &  8.13 &  0.000044 & 1A \\
  C/2000 CT54 &  3.1561  & 10.71 &  0.000051 & 1A \\
  C/2000 H1   &  3.6366  & 10.29 &  \nodata  &    \\ 
  C/2000 J1   &  2.4371  & 12.62 &  0.001406 & 1A \\
  C/2000 O1   &  5.9218  &  7.03 &  0.000037 & 1A \\
  C/2000 OF8  &  2.1731  & 14.07 &  0.000048 & 1B \\
  C/2000 SV74 &  3.5416  &  9.40 &  0.000090 & 1A \\
  C/2000 U5   &  3.4852  &  9.88 &  0.000358 & 1A \\
  C/2000 W1   &  0.3212  & 10.44 &  \nodata  &    \\
  C/2000 WM1  &  0.5553  &  6.81 & -0.000459 & II \\
  C/2000 Y1   &  7.9747  &  9.54 &  0.000063 & 1A \\
  C/2000 Y2   &  2.7687  &  9.65 &  0.001934 & 1B \\
  C/2001 A1   &  2.4062  & 10.71 &  0.005738 & 2A \\
  C/2001 A2   &  0.7790  & 14.22 &  0.000447 & II \\
  C/2001 B1   &  2.9280  & 11.14 &  0.000071 & 1B \\
  C/2001 B2   &  5.3065  &  5.60 &  0.000187 & 1B \\
  C/2001 C1   &  5.1046  & 10.30 &  0.000020 & 1A \\ 
  C/2001 G1   &  8.2356  &  7.45 &  0.000024 & 1A \\
  C/2001 HT50 &  2.7921  &  3.15 &  0.000878 & 1A \\
  C/2001 K3   &  3.0601  &  9.80 &  0.000072 & 1B \\
  C/2001 K5   &  5.1843  &  8.10 &  0.000029 & 1A \\
  C/2001 N2   &  2.6686  &  5.77 &  0.000455 & 1A \\
  C/2001 RX14 &  2.0576  &  6.06 &  0.000776 & 1A \\
  C/2001 S1   &  3.7500  & 11.36 &  0.018168 &    \\
  C/2001 U6   &  4.4064  &  7.42 &  0.000998 & 1A \\
  C/2001 W1   &  2.3995  & 14.04 &  \nodata  &    \\
  C/2001 X1   &  1.6976  & 12.54 &  0.002285 & 2A \\
  C/2002 B2   &  3.8430  & 10.12 &  \nodata  &    \\
  C/2002 B3   &  6.0525  &  7.92 &  \nodata  &    \\
  C/2002 C2   &  3.2538  &  8.88 &  0.000393 & 1B \\
  C/2002 E2   &  1.4664  & 10.34 &  0.000173 & 1B \\
  C/2002 H2   &  1.6348  & 13.29 &  0.004024 & 2A \\
  C/2002 K2   &  5.2378  &  7.62 &  \nodata  &    \\
  C/2002 L9   &  7.0316  &  5.60 &  0.000035 & 2A \\
  C/2002 O4   &  0.7762  & 13.59 & -0.000772 & 2A \\
  C/2002 P1   &  6.5307  &  8.55 &  0.002023 & 2A \\
  C/2002 Q2   &  1.3062  & 16.09 &  \nodata  &    \\
  C/2002 Q5   &  1.2430  & 16.59 &  0.000058 & 1B \\
  C/2002 U2   &  1.2086  & 14.63 &  0.001075 & 1B \\

\enddata

\tablenotetext{a}{Where available, this is the reciprocal of the original semi-major axis, 
ie. before planetary perturbations. Taken from \citet{mar03}}

\tablenotetext{b}{Quality flag for the orbit determination. Original semi-major axes only
available for classes 1A thru 2B.}

\end{deluxetable}

\subsection{Properties of the Sample\label{prop}}

\begin{figure*}
\plotone{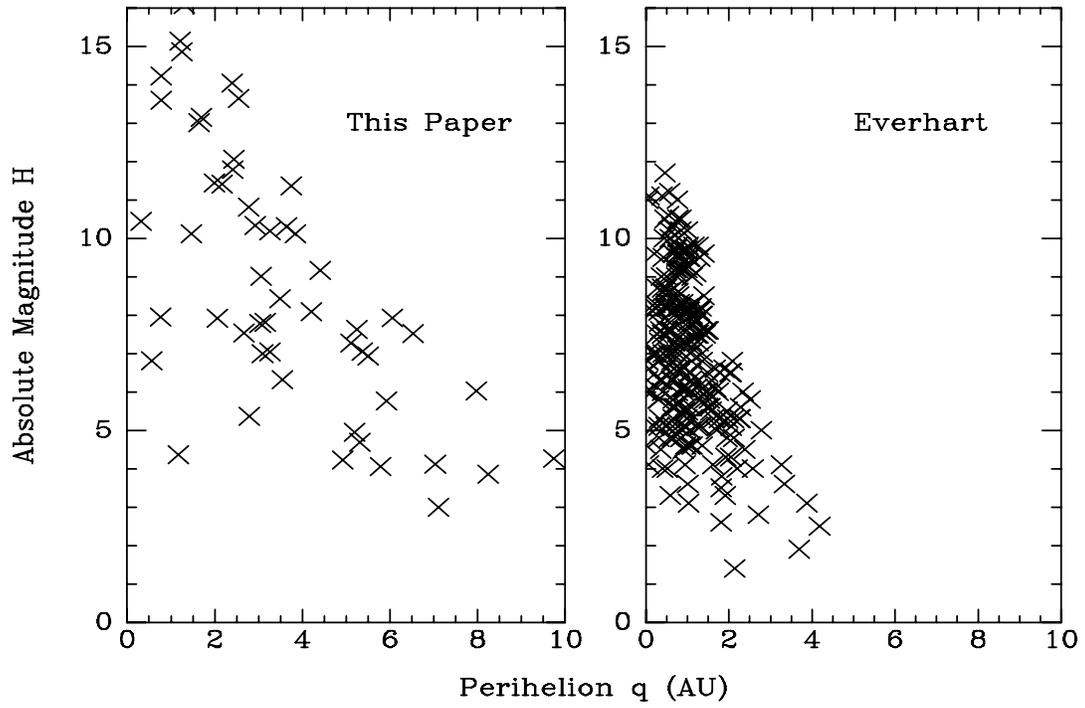}
\caption{
The location of our comet sample in the absolute magnitude vs perihelion plane, as compared to the
Everhart (1967) sample.
\label{everhart_compare}}
\end{figure*}

The LINEAR sample has very different properties from historical samples (as typified by the Everhart
sample). Figure~\ref{everhart_compare} shows that the LINEAR sample extends around 4 magnitudes
deeper, and to much larger perihelia. The overlap is small: only $\sim 5$ of the LINEAR comets lie
within the absolute magnitude and perihelion region sampled by historical samples.

\begin{figure*}
\plotone{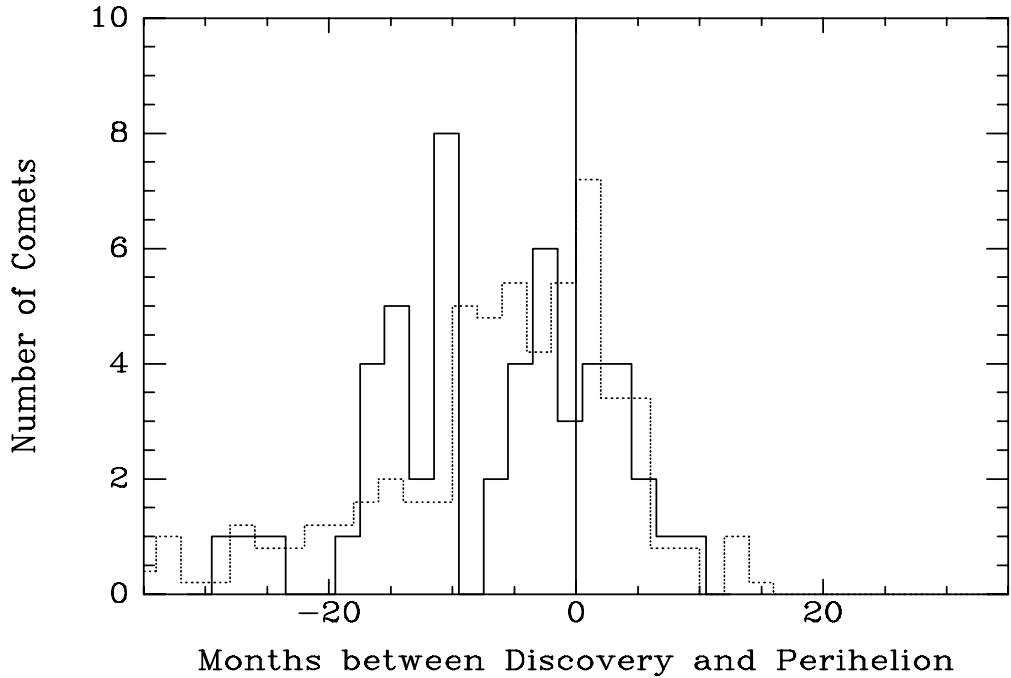}
\caption{
The time interval between the discovery of a comet and its perihelion passage (solid line). 
Negative values indicate that the comet was discovered before passing perihelion. For comparison,
the dotted line shows our prediction of this distribution, from the best-fit model comet
population.
\label{discover_hist}}
\end{figure*}

Analysis of the discovery telegrams indicates that almost all of the comets in the sample were
originally identified as moving point sources. They were posted on the Near Earth Object (NEO) 
confirmation page\footnote{\url{http://cfa-www.harvard.edu/iau/NEO/ToConfirm.html}} at the
Minor Planet Center. Follow-up observations then determined that the sources were spatially extended
and hence comets. 77\% were discovered before reaching perihelion (Fig~\ref{discover_hist}), and 
73\% were were first detected when more than 3AU from the Sun.

The necessity for follow-up potentially introduces two sources of incompleteness into this
sample. Firstly, some fraction of objects posted on the NEO confirmation page are never followed
up in enough detail to determine whether they are comets or not. Timothy Spahr kindly provided 
records of all objects posted to the NEO confirmation page in 2000-2002.
 Only 11\% of these were not followed-up well 
enough to determine an orbit: this places an upper limit on the fractional incompleteness of
our sample due to failed follow-up. This is probably a conservative upper limit: most of these
lost objects were most likely either not real to begin with or fast-moving objects only visible
for a short window of time. Secondly, some comets might have been inactive at these large
heliocentric distances, and hence classified as minor planets. The minor planet 
centre database was checked for non-cometary objects on long-period, highly eccentric orbits, 
but only one was found which reached perihelion within the period 2000-2002:
2002 RN109. Thus this too is not a major source of incompleteness. It also shows that most
comets down to the LINEAR magnitude limit are still active out to 10AU from the Sun.

\begin{figure*}
\plotone{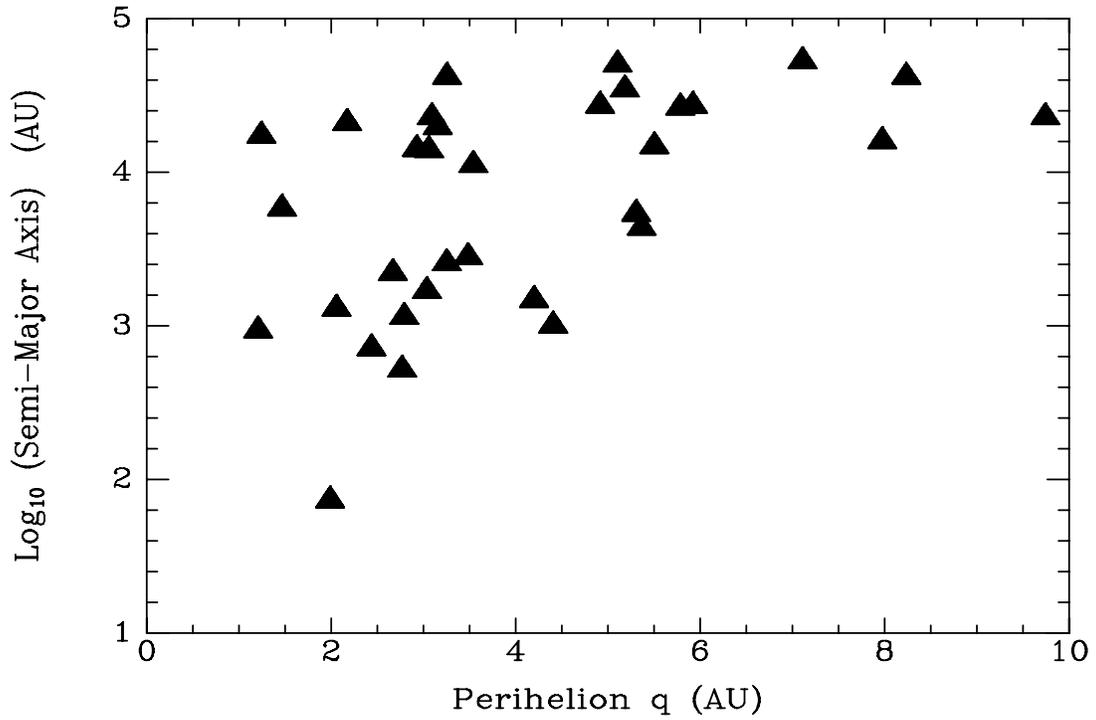}
\caption{
The perihelia and semi-major axes of all comets in the LINEAR sample with class 1A or 1B orbit
determinations. The distribution remains largely unchanged if comets with less well determined
orbits are included.
\label{q_a}}
\end{figure*}

Fig~\ref{q_a} shows an intriguing correlation between perihelion distance and semi-major
axis in the LINEAR sample. This correlation was first noted by \citet{mar73} at smaller 
perihelia. They suggested that
it was a selection effect. Dynamically new comets are brighter at large heliocentric
radii \citep[eg.][]{whi78}, presumably due to extra outgassing at large heliocentric distances 
from their
relatively pristine surfaces, due perhaps to $CO_2$ or a water ice phase transition. The 
Whipple data did not extend to distances beyond 4AU from the
Sun. If this trend continues to larger heliocentric distances, however, it would make dynamically 
new comets far brighter than older comets with the same absolute
magnitude. This could thus, in principle, bias the sample heavily towards new comets, and hence
larger semi-major axes. 

\begin{figure*}
\plotone{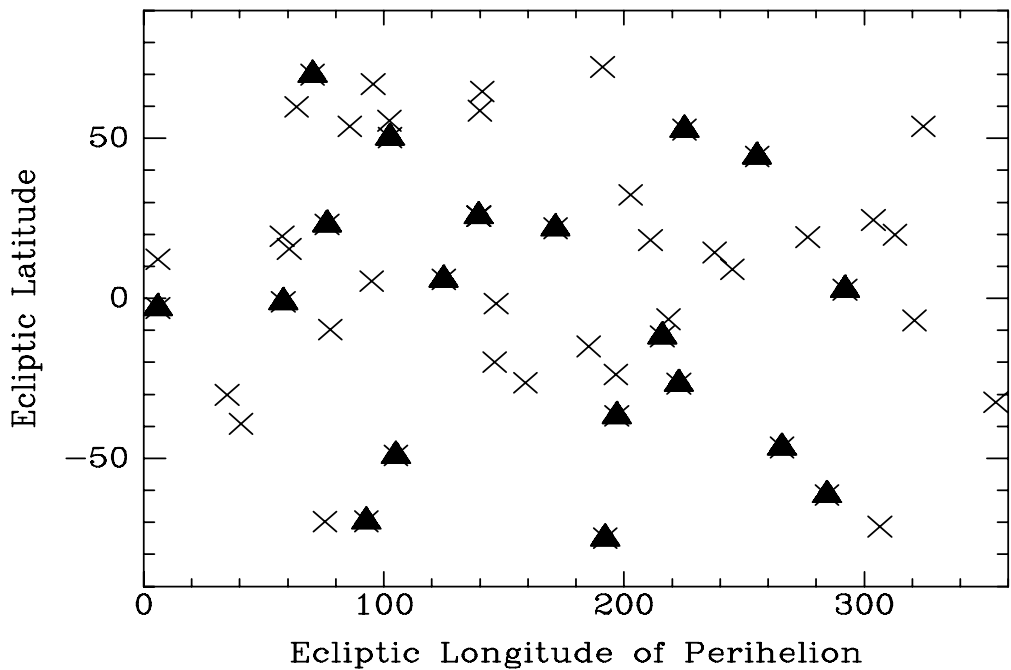}
\caption{
The ecliptic coordinates of the perihelia of comets in the LINEAR sample. Solid triangles are
comets which class 1a or 1b orbits which are dynamically new (as defined in the text).
\label{peripos}}
\end{figure*}

\begin{figure*}
\plotone{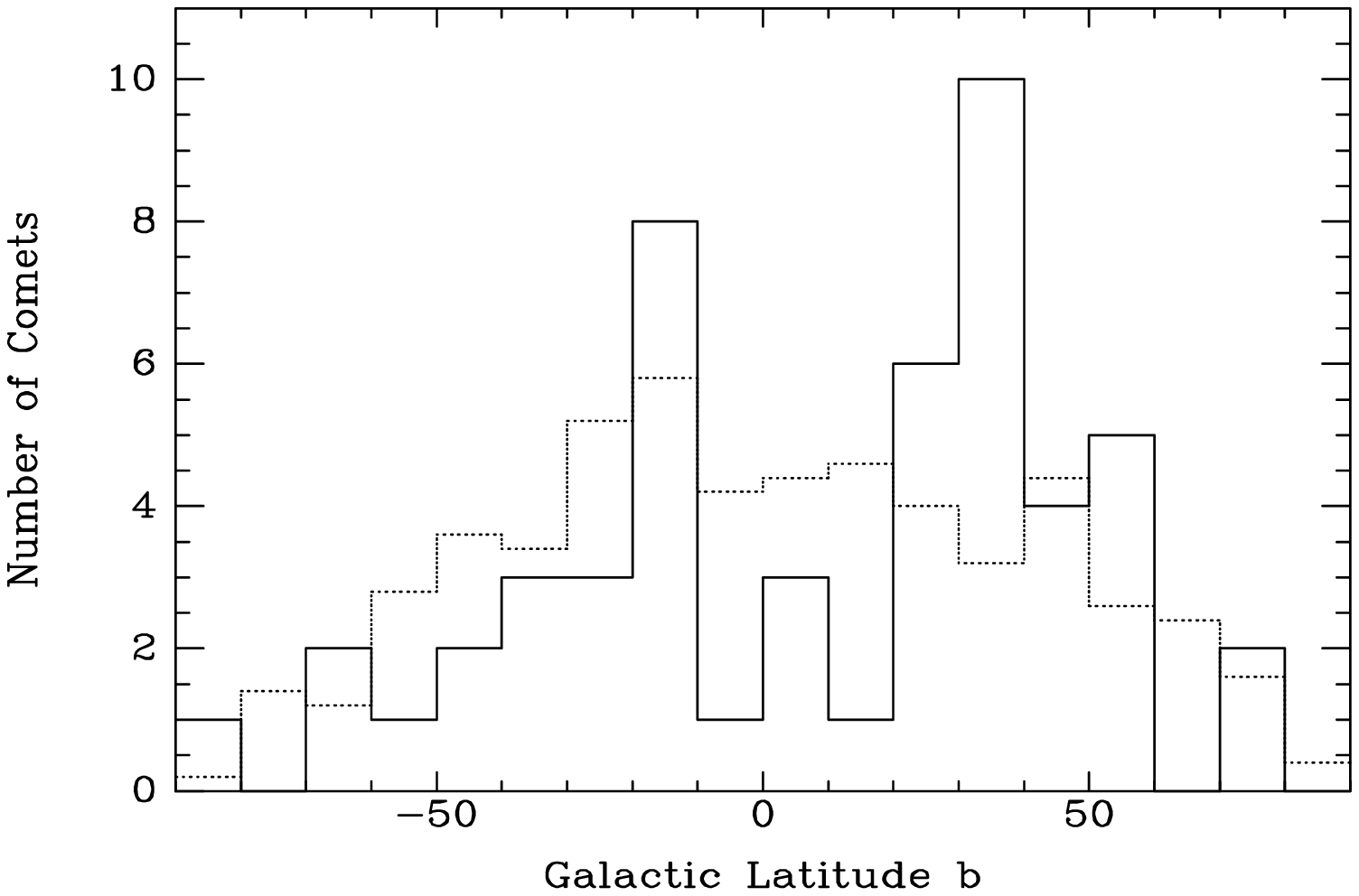}
\caption{
The galactic latitude distribution of the perihelion directions of the LINEAR sample comets
(solid line). The dotted line is the predicted distribution from our best-fit model
(\S~\ref{best}).
\label{bhist}}
\end{figure*}

The distribution of perihelion positions is shown in Fig~\ref{peripos}. The comets are
weakly concentrated at intermediate galactic latitudes (Fig~\ref{bhist}), consistent with
the galactic tide playing a major part in making them observable \citep{mat96}. The galactic
latitude distribution is not, however, significantly different from the predictions of a
best-fit model (\S~\ref{best}) assuming a random distribution, as measured by the 
Kolmogorov-Smirnov (KS) test or the Kuiper statistic. There are also no significant 
great-circle alignments \citep{hor02}.

\section{Model of the Long Period Comet Population}

\label{model}

The LINEAR comet sample was compared against a Monte-Carlo simulation of the long period comet
distribution. In this section I discuss the parameters used in this simulation.

\subsection{Orbital Parameters}

The \citet{eve67} and \citet{hug01} studies are based on comets with a limited range of
perihelion distances q and hence give only weak constraints on this distribution. Everhart
found a factor of two rise in the number of comets per unit perihelion between 0 and 1 AU, but 
beyond that the data are consistent either with a continuing rise or a flat distribution. Hughes
found no significant trend in number of comets against perihelion, but his data are quite consistent
with such a trend.

On theoretical grounds, however, a gentle rise in the number of comets as a function of
perihelion is expected, as comets diffuse into the solar system past the barrier of giant
planet perturbations \citep[eg.][]{tsu92, wei99}. As a first guess, I chose to model the
perihelion distribution as an unbroken straight line:
\begin{equation}
\frac{dn}{dq} = 1 + Aq,
\label{perieq}
\label{qeqn}
\end{equation}
where $A=1$ gives a reasonable fit to the \citet{eve67} distribution.
The \citet{wei99} slope is shallower, but only shown out to 3AU.

The distribution of semi-major axes $a$ makes no significant difference to the conclusions,
as the comets are all very close to being on parabolic orbits. I chose to randomly class 40\% 
of comets as dynamically new and given them all $a=$20,000, while the remainder were given a value
of $a$ randomly and uniformly distributed between 1000 and 20,000 AU.

The time of perihelion passage, orbital inclination and perihelion direction were 
randomly chosen to give a uniform distribution 
on the celestial sphere. This ignores possible great-circle alignments \citep{hor02}
and galactic tidal effects \citep{mat96,mat04}.

\subsection{Absolute Magnitudes\label{absmag}}

\citet{eve67} found that the absolute magnitude distribution of
comets was best fit by a broken power-law, with the break at $H \sim 6$. \citet{hug01} also
found a break at about the same absolute magnitude, but was unable to decide whether it
was a real break or simply the effect of increasingly incomplete samples at fainter magnitudes.
To bracket the possibilities, I use a broken power-law of the form.
\begin{equation}
\frac{dn}{dH} \propto  \left\{ \begin{array}{ll}
                       b^{(H - H_b)}, & \ H < H_b \\
                       f^{(H - H_b)}, & \ H > H_b, \\
                     \end{array}
              \right. 
              \label{heq}
\end{equation}
where $H_b$ is the break magnitude, $b$ is the bright end slope and $f$ is the faint end 
slope. Everhart gives $H_b = 6$, $b=3.65$ and $f=1.82$. Hughes gives $H_b = 6.5$ and
$b=2.2$. If I assume that his observed break is real and not an artefact of sample
incompleteness, his plots imply a faint end slope of $f=1.07$, which I adopt to bracket
the possibilities. This version of the Hughes formulation thus predicts dramatically fewer faint
comets, as would be expected as these are the ones for which Everhart applies the
largest incompleteness correction fraction.

\subsection{Comet Flux}

\citet{eve67} and \citet{hug01} give different estimates of the long-period comet
flux through the inner solar system. Everhart estimates a flux of 8000 comets with $H<10.9$ and
$q < 4$ over 127 years. Hughes estimates a flux of  
0.53 comets per year brighter than $H = 6.5$ per unit perihelion. I ran the
simulations using both.

\section{Monte-Carlo Simulation}

\label{montecarlo}

For a given model comet population, the aim is to simulate the observable properties of a
sample that matches the selection effects of the LINEAR sample.
The simulation starts off by generating a set of comets that reach perihelion within a
three year period. The comets are randomly generated using the model distributions in the
previous section. The model extends down to $H=19$ and out to $q=15$. 

I generated two model populations: one using the Everhart absolute magnitude distribution and
flux, the other using the Hughes absolute magnitude distribution and flux. In the Everhart model
260,000 comets are generated, while only 4,000 are needed in the Hughes model.

The position of each comet is then calculated at 24 hour intervals throughout the three year
period, and its heliocentric distance $r$, distance from the Earth $\Delta$, apparent celestial
coordinates and apparent angular velocity written to file. Pure elliptical orbits are used:
no attempt is made to allow for planetary perturbations.

At each location, the apparent total magnitude is then calculated. Comets are notoriously variable
in how rapidly their apparent magnitude varies as a function of heliocentric distance. I parameterize 
this, as is conventional, using Equation~\ref{eqabsmag}. 
Two values of $n$ are randomly assigned to each comet: one for before perihelion and another
for after. For the 40\% of our simulated comets that I set as dynamically new, the
pre-perihelion value of $n$ is chosen from a Gaussian distribution of mean 2.44 and
standard deviation 0.3. Post-perihelion, the mean is 3.35 with a scatter of 0.27. For the
remaining comets, the pre-perihelion numbers are 5.0 with a scatter of 0.8, and after perihelion
3.5 with a scatter of 0.5. All these values are taken from \citet{whi78}. At large distances
from the Sun, cometary activity will presumably stop, and a bare nucleus will have $n=2$.
The near-ubiquitous detection of fuzz around the LINEAR comets implies, however, that this only 
happens further from the Sun than our models reach.

This approach can only be a rough approximation to the real radial brightness dependence.
The value of $n$ for an individual comet is typically time dependent, and all the tabulated values
are for comets within $\sim 3$ AU of the Sun, whereas our simulation tracks them out beyond
10AU. In addition, comets show occasional flares above and beyond this power-law behavior, which I
have not attempted to model. Such flares might introduce an amplification bias, with comets being
pushed over the detection threshold. As we will see, however, the slope of the
absolute magnitude distribution is so gentle that this is unlikely to be a major effect.

\subsection{Converting total magnitudes to point-source equivalent magnitudes\label{mags}}

The apparent total magnitude of each simulated comet can now be calculated at any given point
in its orbit. Unfortunately, in any CCD-based survey, it is the peak surface brightness of the
coma that determines whether something has been seen, not the total magnitude. The LINEAR
skymaps, furthermore, list only the magnitude limit {\em for a point source} at any given location on the sky 
(typically around 19).

As discussed in the introduction, total cometary magnitudes are notoriously unreliable. 
Quantitative studies prefer more reproducible and
physically meaningful parameters such as $Af\rho$ \citep[eg.][]{ahe95}. Unfortunately, not enough 
long period 
comets have been studied in this way to derive the $Af \rho$ distribution. We are therefore
forced to attempt some conversion between total magnitudes and point-source equivalent magnitudes.

\begin{figure*}
\plotone{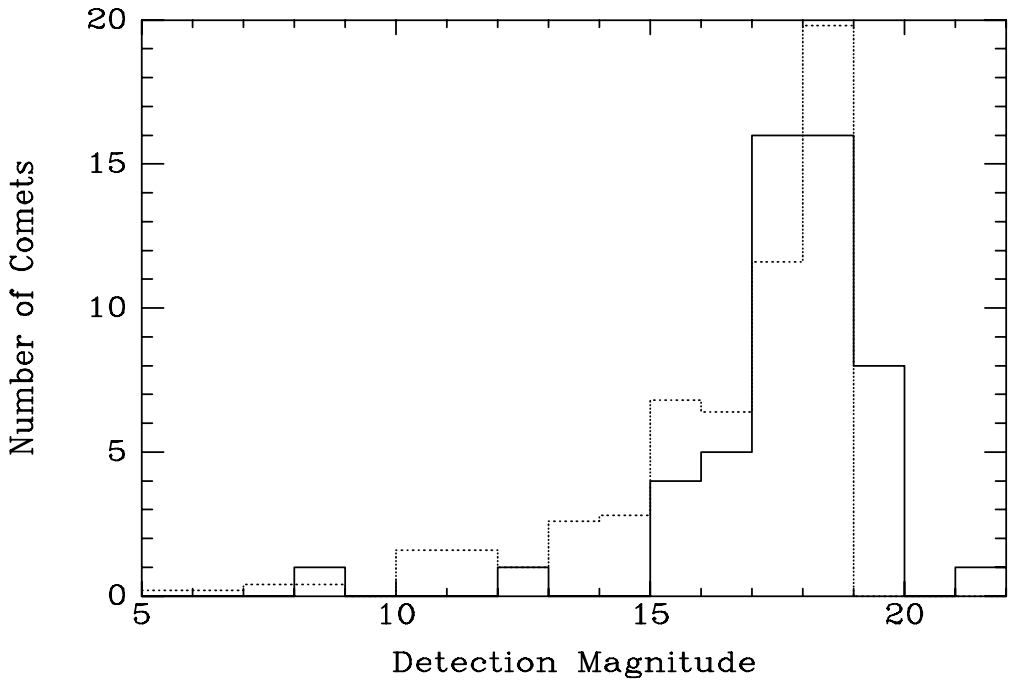}
\caption{
The predicted (dotted) and observed (solid) distributions of discovery magnitudes.
\label{maghist}}
\end{figure*}

For bright and near-by comets, this correction can be as large as $\sim 5$ magnitudes 
\citep[eg.][]{fer99}. The comets in the LINEAR 
sample were, however, typically first seen when very faint (Fig~\ref{maghist}), and were generally 
mistaken for point sources in the initial observation. We might therefore expect the correction 
factor to be much smaller, 
at least when close to the detection threshold.

The histogram of detection magnitudes (Fig~\ref{maghist}) climbs steeply down to $H_{\rm dis} 
\sim 19$, and
then falls off fast (the one comet discovered when fainter than 20th mag was found by
Spacewatch, which has a fainter magnitude limit). This fall-off occurs at almost exactly the
same magnitudes as LINEAR's point source limit, which ranged from around 18 to 20. I thus conclude 
that near the LINEAR detection threshold, total magnitudes and point source equivalent
magnitudes are similar. When generating mock samples, it is only the magnitude near the
detection threshold that determines whether or not a given model comet is included in the mock catalog.

The exact value of this correction value was set iteratively. I initially guess that the point-source 
equivalent (PSE) magnitude and total magnitude (TM) are the same, and run the
simulations of the comet sample. I use the model that best fits the data (\S~\ref{best}) to calculate
the predicted discovery magnitude distribution, and compare this to the observed
distribution. I then tweak the PSE$-$TM correction to bring the histograms into agreement.
The best match is obtained when PSE$-$TM$= 0.5 \pm 0.5$ (Fig~\ref{maghist}). I use a value 
of $0.5$ throughout this
paper, except where otherwise noted.

The predicted magnitude is corrected
for the effects of trailing. LINEAR exposure times vary from 3 to 12 sec: the latter
was used in the correction as it minimised the predicted number of very faint comets. 2\arcsec\ 
seeing (FWHM) was assumed. Trailing makes very little
difference, except for the very faintest comets. LINEAR uses unfiltered CCD magnitudes while
the historical surveys use unfiltered visual magnitudes.  These will be somewhat
different, due to the different wavelength sensitivity of the human eye and of the LINEAR
CCDs, but the discrepancy should only be a few tenths of a magnitude at most, and hence is not a 
dominant source of error.

Anoher possible worry: the absolute magnitudes I quote for the comets in the LINEAR sample 
(Table~\ref{sampletab})
are derived from total magnitudes measured when the comets were barely resolved and far 
from the Sun, using a model for the heliocentric brightness variation. The absolute
magnitudes fit by Everhart and Hughes are based on observations of highly extended comets observed
close to the Sun and Earth. These are thus very different quantities, and might well 
be systematically different, if there is some error in our heliocentric brightness correction, 
if total magnitudes for barely resolved comets are systematicaly different from total 
magnitudes for greatly extended comets, or if there is some systematic bias in the discovery 
magnitudes reported to the central bureau of astronomical telegrams.

To test this, I picked out the five comets in the LINEAR sample which were discovered when
far (more than 3.5 AU) from the Sun, but which subsequently passed close enough to the Earth 
and Sun for traditional small telescope visual magnitude estimates (within 2AU). 
Dan Green kindly provided me with compilations of visual magnitude estimates of these
comets while they were close to the Earth and Sun, taken from the archives of the
International Comet Quarterly\footnote{\url{http://cfa-www.harvard.edu/icq/icq.html}}. These
visual/small telescope magnitude estimates should be broadly comparable to the data on which
the Everhart and Hughes papers were based. I then compared the predicted magnitudes when
close to the sun (based on the discovery magnitude and the model in this paper) with the 
tabulated observations. There was a considerable scatter in the measured visual magnitudes for each 
comet: I simply averaged all visual small telescope magnitudes made when the comet was as
close as possible to 1AU from both Sun and Earth.

My predicted magnitudes were consistent with the observed values, albeit with a large scatter.
The mean difference (predicted magnitude minus observed magnitude) was $0.4 \pm 0.7$, where the
error indicates the $1 \sigma$ dispersion of the mean. This
is not, alas, a strong constraint, but does indicate that the two magnitude scales are not
grossly discrepant.

\subsection{LINEAR's Sky Coverage}

The final step is to determine whether LINEAR imaged a part of the sky in which the comet
was detectable and within its magnitude limit. 
The published LINEAR skymaps show that during each dark period in 2000-2002, they attempted to survey
the region whose midnight hour angle is in the range $-7 < {\rm HA} < 7$, and in the
declination range $-30 < \delta < +80^{\circ}$. In winter months with good weather, they surveyed
more than 90\% of this whole region down to a point-source magnitude limit of
better than 19. In bad months, this dropped to a magnitude limit of around 18.5 over 60\% of
this region, and occasionally worse. In the mean month, 72\% of this region was surveyed
to a visual magnitude limit of 18.5 or better. The exact pattern surveyed was complex and
variable: the only constant was that the densest regions of the galactic plane were avoided.
Each field was imaged five times in succession, with 3 -- 12 sec per exposure, once in 
every dark period.

This sky coverage was approximated as follows.
Each comet that enters the $-7 < {\rm HA} < 7$, $-30 < \delta < +80^{\circ}$ region at any
point, with a point-source equivalent (PSE) magnitude brighter than 19 is considered to have
been potentially observable, unless it was within ten degrees of the galactic plane.
If a comet is predicted to be detectable for an entire lunation, it is given a 80\% chance
of having been detected during that lunation. If it was predicted to be visible for less than
the whole lunation, it is given a probability of having been detected equal to 80\% of the
fraction of the lunation for which is was potentially observable.

Does this approximation match the real, more complex selection function? This was tested by 
manually checking 100 simulated comet ephemerides, containing monthly positions and magnitudes, 
against the real LINEAR sky-maps. The
approximation was found to give a number and absolute magnitude distribution of detected comets
indistinguishable from the manually checked sample.

This approach should slightly overestimate the probability of a comet being observed, as comets 
could be blended with star or galaxy images. Experience suggests that at this relatively bright
magnitude limit, this is only a few percent effect at worst, at least away from the denser
regions of the galactic plane, which the survey did not cover.

Another possible source of error is sky subtraction. It is unclear exactly how the LINEAR
survey do their sky subtraction, but if some of the extended coma emission is included in the
sky value, this will artificially suppress the point source equivalent magnitude. The worst
case sky subtraction algorithm would be to measure the sky brightness from an annulus close to
the comet nucleus. If I assume that the sky is measured only 5\arcsec\ from the nucleus (unlikely), 
we can used the observed $1/r$ surface brightness profiles of cometary comae \citep{jew87} to show
that this sky subtraction algorithm would reduce the measured comet brightness by $\sim 0.2$ mag.
More plausible sky subtraction schemes would reduce it by less, or not at all. Thus this too is
not a dominant source of error.

\section{Comparison with the Models}

\label{compmodels}

In this section, I compare the data to the Monte-Carlo simulations of what LINEAR would
have seen over a three year period. In the scatter plots, the data are compared to a single run of
the simulation. In all histograms and quoted statistics, however, the data are compared against 
the average or
sum (as appropriate) of five Monte-Carlo runs based on the same comet population model. This
summation should suppress the error due to small number statistics in the simulated samples to 
well below that of the observed sample.

\subsection{Everhart\label{eve}}

The observed sample properties are first compared against the Monte-Carlo prediction
using the \citet{eve67} flux normalization and absolute magnitude distribution.

\begin{figure*}
\plotone{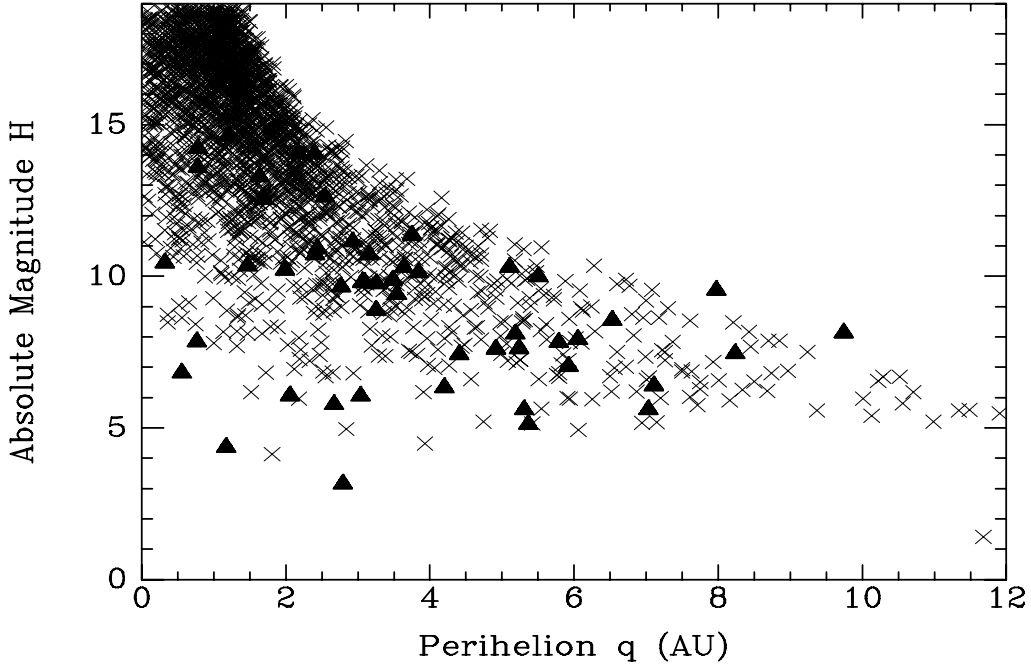}
\caption{
The perihelia and absolute magnitudes of model comets (crosses) and the LINEAR sample
(triangles). The model comet population was generated using the \citet{eve67} flux and 
absolute magnitude distribution.
\label{complot_eve}}
\end{figure*}

\begin{figure*}
\plotone{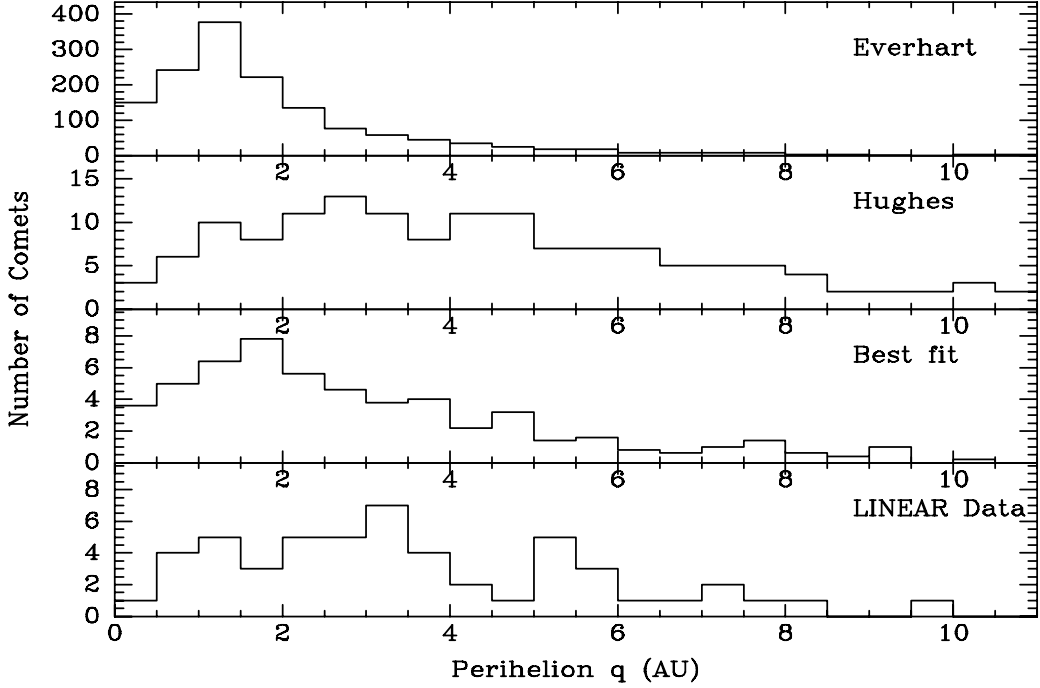}
\caption{
The perihelia distributions predicted by three different model comet
populations, compared to the observed distribution for the LINEAR
sample.
\label{qhist}}
\end{figure*}

\begin{figure*}
\plotone{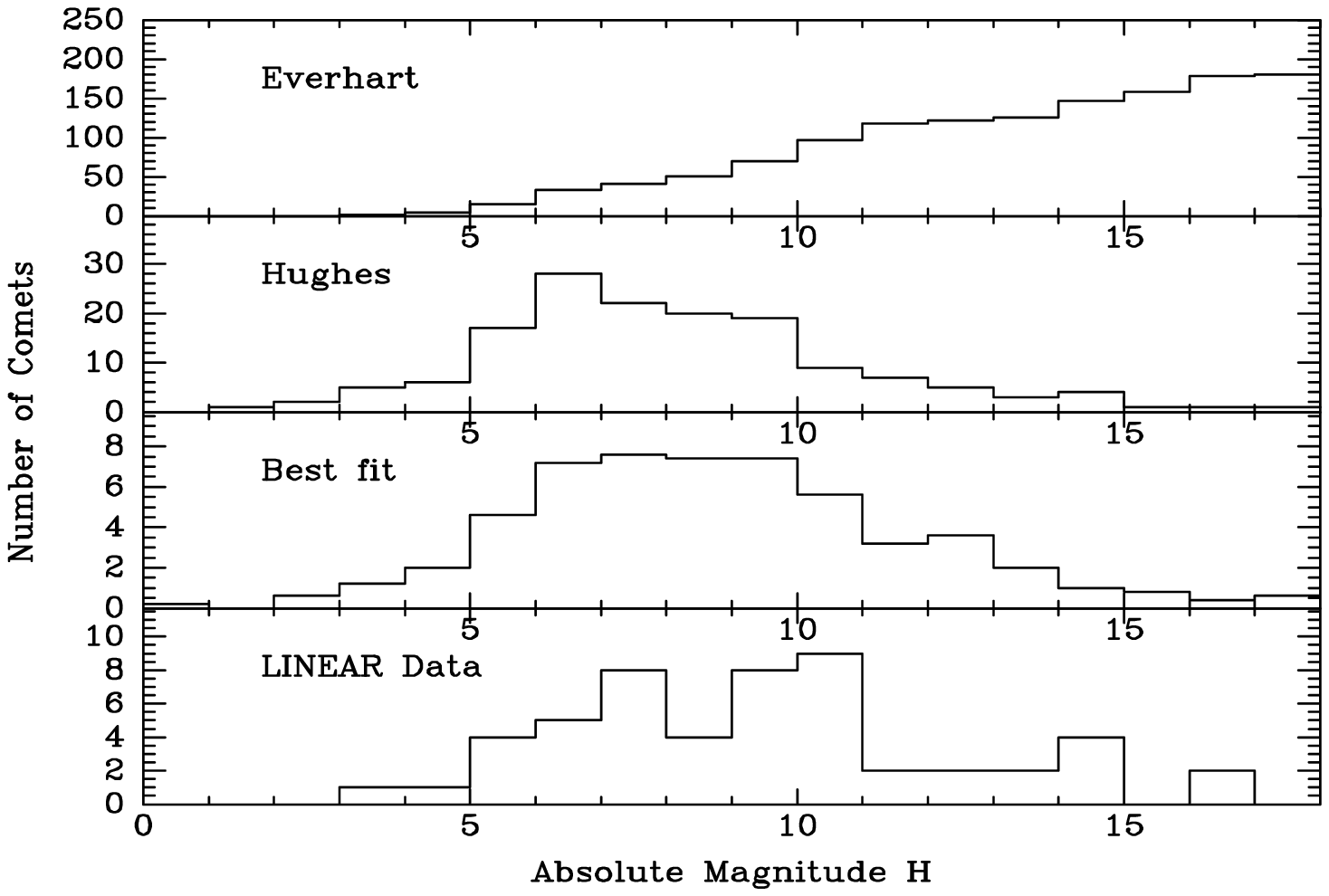}
\caption{
The absolute magnitude distributions predicted by three different model comet
populations, compared to the observed distribution for the LINEAR
sample.
\label{hhist}}
\end{figure*}

Figs~\ref{complot_eve}, \ref{qhist} and \ref{hhist} compare the distribution of model and 
observed comets in perihelion
$q$ and absolute magnitude $H$. The upper boundary to the locus of points is set by the
magnitude limit, and seems a reasonable fit to the data. But the model clearly predicts far
too many comets: 2228 as compared to the 52 observed.

The discrepancy is primarily at fainter absolute magnitudes: brighter than $H \sim 7$ the model
and data are consistent. The worst discrepancy is for comets fainter than $H \sim 11$: ie. fainter than 
the data on which Everhart based his model. It is thus a test of the power-law extrapolation.
The model predicts that LINEAR should have seen 1848 comets fainter than $H = 11$, whereas
only 12 were seen. Irrespective of the flux normalization, the shape of the absolute magnitude 
distribution (Fig~\ref{hhist}) is wrong: a KS-test comparison with the observed distribution 
shows that they are
inconsistent with $>99.99\%$ confidence.

Is this discrepancy real, or is there some reason why LINEAR would miss faint comets close to the
Earth? The model comets with $H>11$ are predicted to be observable for a median period of 52
days, at a median distance from the Earth of $\Delta = 1.0$ AU. Their typical apparent
angular velocity is predicted to be $\sim 1.1$ degrees per day. Their observational properties
are thus typical of near Earth objects, which LINEAR finds in profusion. It is thus hard to see
what selection effect could prevent their detection. Grant Stokes (personal communication) confirms
that there is nothing in their data analysis which should preclude the discovery of comets
like these. Image trailing and short observability windows do reduce the number of these
comets seen, but these are already taken into account by the Monte-Carlo simulation.

Could the discrepancy be an artifact of the various approximations made in the model? The
discrepancy is insensitive to assumptions about the dependence of brightness or comet number 
on heliocentric
distance, as these sources are observed at close to 1AU. One possibility is that I have 
incorrectly estimated the effective magnitude limit of LINEAR for sources with these apparent
total magnitudes. To test this, I repeated the analysis with a magnitude limit set two magnitudes
brighter than my best estimate. This reduced the discrepancy but did not remove it: the prediction 
dropped to 577 observed comets fainter than $H=11$, still more than two orders of magnitude above
the data. None of the plausible incompletenesses in the data, nor other assumptions in the
model can come close to removing this discrepancy. I therefore conclude that the Everhart model
cannot be extrapolated to absolute magnitudes fainter than $H=11$.

Even at brighter absolute magnitudes, however, there remains a substantial discrepancy. The
model predicts that 188 comets with $H<11$ and $q<4$ should have entered the solar system
within the three year period, and that 86\% of them (162) would have been detected by LINEAR.
Only 21 such comets were observed. It is hard to see that LINEAR could have missed many comets
this bright passing this close to the Sun. The model predicts that these comets should 
remain visible for a median 208 days (7 lunations), so almost regardless of position on the sky, 
they should have had several opportunities to be observed. They spend much of this time many 
magnitudes above the survey detection limit. Indeed, the 21 comets observed with these 
properties were discovered a median 11 months before perihelion, at a median heliocentric 
distance of 4.3AU, confirming that these are easy targets. The discrepancy occurs mostly at 
the fainter magnitudes within this range: brighter than $H \sim 7$ there is
no significant difference between the Everhart predictions and the LINEAR observations.

I conclude, therefore, that the Everhart model fails in two ways. Firstly, the quoted 
normalization of 8,000 comets per 127 years with $H<10.9$ and $q<4$ is too high by a factor of
$\sim 7$. Secondly, the faint end slope of the Everhart absolute magnitude relation is
much too steep, and immensely over-predicts the number of faint comets. This second conclusion was
first reached by \citet{sek86}: the current paper independently confirms their result.

Both discrepancies suggest that Everhart overestimated the incompleteness of his sample of
long-period comets. The discrepancy goes away where the incompleteness correction is 
small, but is largest at the faint magnitudes where the correction is large. A discrepancy here is
perhaps not surprising, as the correction factors calculated by Everhart were so large:
he corrected the 256 observed comets to a flux of 8000: a factor of 31. My analysis reduces
this correction factor to only $\sim 4$.

One possible reason for the difference: Everhart calculated the detection threshold for
typical historical comet searchers, and assumed that the same threshold applied when searching
for comets initially, and when making follow-up observations of known comets. His model was
validated by noting that the last observations of comets occurred close to the time when his
model suggested that they dropped below detectability. But let us hypothesize that comets just
above the detection threshold might be missed as the telescope speeds past during a scan for
new comets, even though they could be detected when looking hard for an already known comet
at a known position. This would reduce the length of time over which a given comet could
have been detected. Detections would thus have occurred earlier in the detectability
window, and the correction factor for incompleteness would thus decrease.

\subsection{Hughes Model\label{hug}}

\begin{figure*}
\plotone{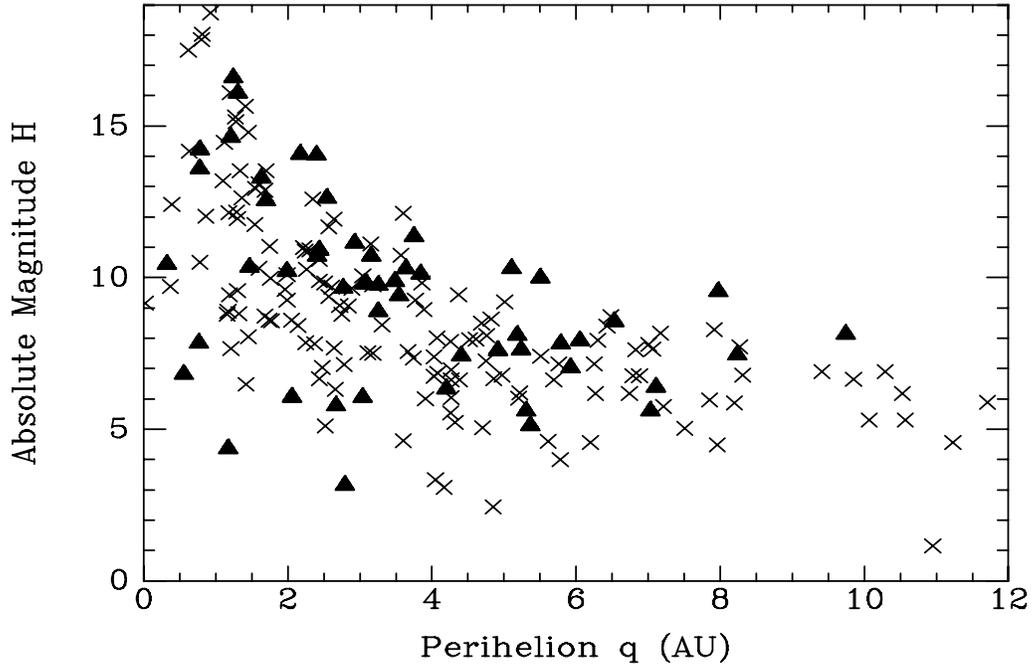}
\caption{
The perihelia and absolute magnitudes of model comets (crosses) and the LINEAR sample
(triangles). The model comet population was generated using the Hughes flux and 
absolute magnitude distribution.
\label{complot_hug}}
\end{figure*}

I now compare the data against a Monte-Carlo simulation using my version of the \citet{hug01}
absolute magnitude distribution and normalization,  combined with
the linearly rising perihelion distribution (Figs~\ref{qhist}, 
\ref{hhist}, \ref{complot_hug}). Once again, the
overall envelope of points agrees well with the model, giving some confidence that the
selection effects have been modeled correctly.

The predictions from the Hughes model are in much better agreement with the data. There is
no vast excess of faint predicted comets, implying that the break seen in Hughes'
data was real, and hence that the flatter faint-end slope of
the absolute magnitude distribution is more accurate. Neither the perihelion nor absolute magnitude
distributions, however, are formally consistent with the observations at the 99\% confidence 
level, as measured by the KS-test. 
The overall flux normalization is also too high: the model predicts that LINEAR should have 
seen 171 comets, rather than
the 52 observed. There is no significant discrepancy within the region in which Hughes
quoted his flux normalization (0.54 comets per year with $H<6.5$ per unit perihelion): the discrepancy
is at fainter absolute magnitudes and larger perihelia. Once again the discrepancy is fairly robust 
against the exact detection limit: dropping the detection threshold by a magnitude reduces
the predicted comet numbers to 148 - still too high.

How can this discrepancy be addressed? Possible incompletenesses in the
LINEAR sample were discussed in \S~\ref{prop} and they can at best increase the observed numbers by
$\sim 20$\%. The Hughes flux normalization is unlikely to be too low, as it was based on
observed counts of very bright comets and made no correction for incompleteness. I therefore
tried to improve the match by tinkering with the absolute magnitude and perihelion distributions.

\subsection{The Best-Fit Model\label{best}}

I first tried reducing the faint end slope of the absolute magnitude relation. If we assume that the
LINEAR sample is 20\% incomplete, we need to reduce the faint end slope from 1.07 down to
0.8 to bring the number of predicted comets down to the observed number. Unfortunately,
this changes the observed absolute magnitude distribution too much: a KS test shows that a model
with this slope predicts an observed $H$ distribution inconsistent with the data with greater 
than 99.99\% confidence.

I then tried changing the perihelion distribution. Decreasing the slope $A$ in 
Eqn~\ref{perieq} to zero brought the predicted number of comets down to 95, but the perihelion
distribution is now too skewed towards small values of $q$ (99.96\% confidence).

I next tried combining both approaches. Decreasing the faint end slope to 1.0, combined
with a flat perihelion distribution, brought the predicted numbers into line with the
observed numbers. Both the perihelion and absolute magnitude distributions were individually
marginally acceptable (KS-test gave 8 and 5\% probabilities of them coming from the
same population as the data) but the joint probability was still uncomfortably low (though the
magnitude limit cut means that the two distributions are not independent, so this should
not be taken too seriously). 
The model predicted too many comets with $q<1$ and $q>5$, and too few in the middle.

Moving the location of the break in the absolute magnitude relation to fainter magnitudes
was also a failure: given that the comet flux normalisation is at brighter magnitudes, this
simply increased the number of fainter comets still further above the observations.

I therefore adopted a different perihelion distribution: one that rises from
$q=0$ out to $q=2$, and is a power-law beyond that. This preserves the Everhart observation
of a drop in comet numbers below $q=1$, while allowing us to tinker with the distribution
further out. The parameterization used was:
\begin{equation}
\frac{dn}{dq} \propto  \left\{ \begin{array}{ll}
                       1 + \sqrt{q}, & \ q < 2 {\rm AU} \\
                       2.41 \times (q/2)^{\gamma}, & \ q > 2 {\rm AU} \\
                     \end{array}
              \right. 
\end{equation}
where $\gamma$ controls the behavior at large perihelia.

\begin{figure*}
\plotone{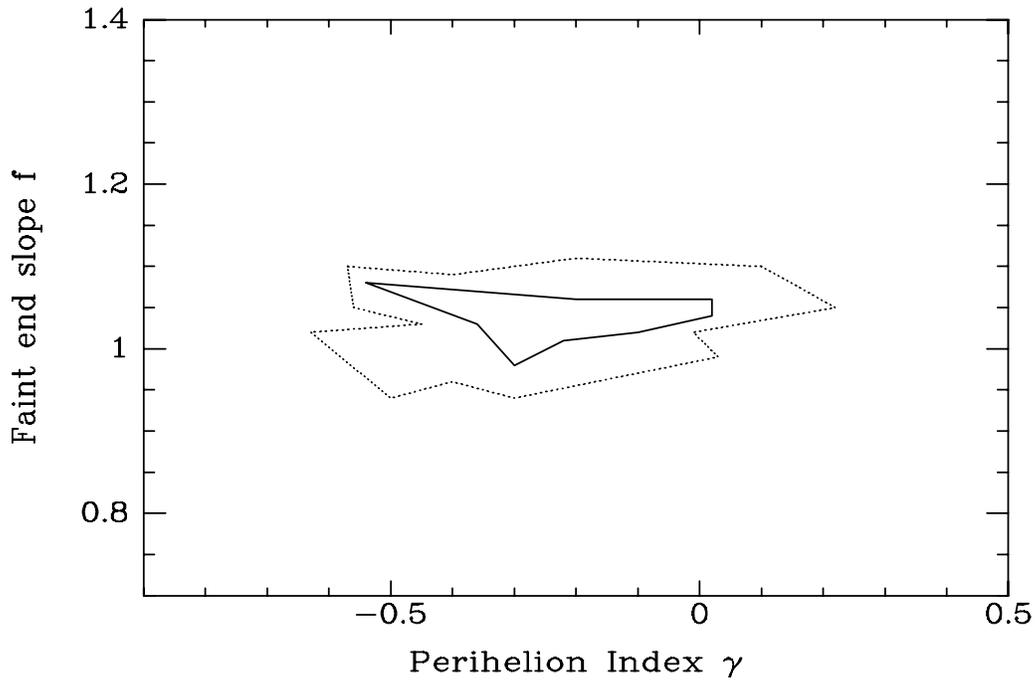}
\caption{
Goodness-of-fit contours as a function of model parameters $\gamma$ and $f$. The solid line is the
95\% confidence contour, and the dotted line is the 99\% contour.
\label{contour}}
\end{figure*}

I then ran a grid of models, varying $\gamma$ and the faint end slope of the absolute
magnitude distribution ($f$ in Equation~\ref{heq}). Each simulated population was tested
against the data in three ways: a KS-test on the perihelion distribution, a KS-test on the
absolute magnitude distribution, and a Chi squared test on the overall predicted number of comets.
The latter was done for both the observed number of comets and a number 10\% higher, to allow for
possible incompletenesses. The lowest of these three significance values was used to
compute the
goodness-of-fit conutours in Fig~\ref{contour}. Note that these contours include only random errors:
the systematic errors are almost certainly larger, especially on $\gamma$. The lowest probability 
rather than the joint probability was used becuase the three tests are not strictly independent.

Quite a tight constraint could be placed on  $f$: $f = 1.03 \pm 0.09$ (95\% confidence, not 
including systematic errors). The constraint on $\gamma$ was weaker: $\gamma = -0.27 \pm 0.3$. No
useful constraint could be placed on the bright-end slope $b$: in the modeling I 
use the \citet{hug01} value of 2.2, but it makes little difference. Note that these slopes are
often described in the literature using the $\alpha$ parameter: $\alpha = \log_{10}{\rm (slope)}$,
so our best-fit faint end slope has $\alpha = 0.004$.

\begin{figure*}
\plotone{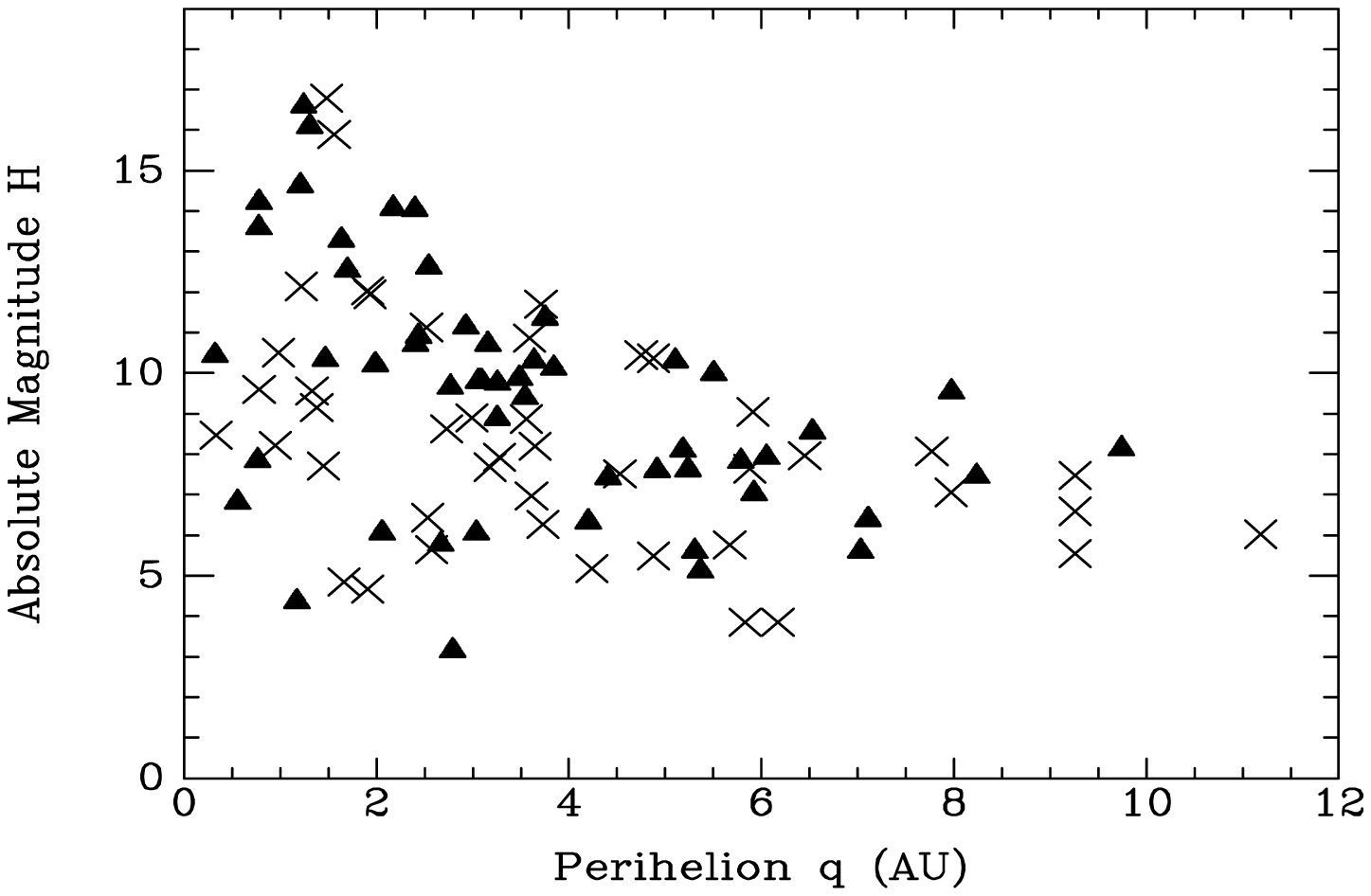}
\caption{
The perihelia and absolute magnitudes of model comets (crosses) and the LINEAR sample
(triangles). The model comet population was generated using our best fit model of the
comet population.
\label{complot_f}}
\end{figure*}

The predicted distribution of comets for the best fit model is shown in Figs~\ref{qhist}, 
\ref{hhist} and \ref{complot_f}.
Models generated using this model predict $55 \pm 3$ ($1 \sigma$) observed comets, in 
excellent agreement with the data.

This model was used to predict the discovery magnitude distribution (Fig~\ref{maghist}).
This comparison was used to set the equivalent point source vs. total magnitude offset, as
described in \S~\ref{mags}. 

My preferred model thus approximately preserves the faint-end slope and normalization derived by 
Hughes. A flat
perihelion distribution is marginally ruled out, and best fits are obtained for one that rises 
from the Sun out to $q \sim 2$ and is either flat or gently falling beyond that.

\section{Discussion}

\label{discussion}

\subsection{The Comet Flux\label{obflux}}

The best-fit model can be used to estimate the flux of long-period comets through the
inner solar system. By definition, this model uses the \citet{hug01} flux of 0.53
comets per year per unit perihelion, brighter than $H=6.5$.

\citet{eve67} estimated a flux of 8000 long-period comets per 127 years with $H<10.9$ and $q<4$. 
My model suggests that the flux of comets with these parameters is much lower:
11 long-period comets per year (1600 per 127 years). The model suggests that LINEAR is
picking up over 60\% of comets with these parameters.

I estimate a true flux of 37 comets/year with $H<15$ and $q<8$, of which LINEAR is
detecting $\sim 40$\%.

\subsection{The Number of Oort Cloud Comets\label{number}}

Many published estimates of the number of comets in the Oort cloud use the \citet{eve67}
comet flux as their starting point, and hence should be revised down. Estimates include
\citet{bai88}, \citet{hei90}, \citet{wei96} and \citet{don04}.

The LINEAR sample includes 22 comets with $H<10.9$ and $q<4$ over a three year period. 5 of these 
were dynamically new, 11 dynamically old, and 6 had orbit determinations too poor to tell. If we 
assume that the
comets without good orbit determinations break up between new and old in the same ratio as the
other comets, we find a flux of 7 dynamically new comets over the three years. My model suggests 
that the LINEAR
sample is $\sim 80$\% complete for finding comets in this range, implying a total flux of
dynamically new comets of $\sim 3$ per year. This corresponds to $\sim 0.8$ per unit perihelion 
per year if a uniform perihelion distribution is assumed. This is a factor of $\sim 2$ lower
than was assumed by \citet{hei90} and \citet{don04}. \citet{bai88} and \citet{wei96} assume
long-period comet fluxes (not just the dynamically new ones) of $\sim 10$ per year per AU down
to $H<10.9$, compared to my value of $\sim 2.5$. After correction, all these estimates come out
roughly the same: 1 -- 3 $\times 10^{11}$ in the outer Oort cloud, down to $H<10.9$. If this is
extended to fainter absolute magnitudes, I estimate an outer Oort cloud population of
$\sim 5 \times 10^{11}$ comets down to $H=17$.

\subsection{The Mass of Oort Cloud Comets}

My model has a much shallower slope of the absolute magnitude distribution than that
of \citet{eve67}, so the average mass of a comet {\em increases}. This cancels out the
decreased number of Oort cloud comets I predict to give a similar total Oort cloud mass to previous
estimates \citep[eg.][]{wei96,don04}, both of which used a mean mass computed by
integrating the Everhart curve.

I used two suggested mass-brightness relations to estimate masses from the observed absolute
magnitudes: one from \citet{bai88} and one from \citet{wei96}. Note that this relation
is extremely
uncertain - very few long-period comets have had even their nuclear magnitudes 
measured. 

The average mass is crucially dependent on the slope of the bright end of the
absolute magnitude distribution, which the data in this paper do not constrain. I bracket 
the possibilities
by using both the Everhart and Hughes values (3.54 and 2.2 respectively). I use my own estimate
of the faint-end slope.

For the Everhart bright-end slope, the total mass converges as you go to brighter magnitudes: the
bulk of the mass resides in comets with $0 < H < 9$. For the Hughes bright-end slope, however,
the total mass diverges as you count brighter comets. The bright comets are rare, but their mass
goes up faster than their number goes down at bright magnitudes. Thus for the Weissman mass
relation, comets with $H \sim -5 $ are 10,000 times more massive than those with $H \sim 5$. The
Hughes bright-end slope, however, suggests that they are only 2,600 times less common, so the total
mass in the brighter comets is actually four times greater. For the Everhart bright-end slope, however,
the brighter comets would be 300,000 times less common.

The Everhart slope is thus physically more appealing, as it avoids the need for a bright cut-off.
Given the success of both the Hughes normalization and faint-end slope in fitting our sample, 
however, his bright-end slope should perhaps be taken seriously, leading to the prospect of an
Oort cloud dominated (in mass terms) by very large comets. In this section, I will cut off the
magnitude range of comets at $H=-5$, but this is arbitrary and will have a large effect on the 
total Oort cloud mass if the Hughes distribution is assumed. \citet{ber04} showed that Kuiper Belt 
objects have a break in their mass distribution at sizes of $\sim 100$ km, which corresponds to 
$H \sim -3$ (\S~\ref{small}). This may or may not apply to Oort cloud comets.

If I take the Everhart slope, the mean mass of comets down to $H=11$ is $1.7 \times 10^{17}$g 
for the
\citet{wei96} mass relation, and $5.6 \times 10^{16}$g for the \citet{bai88} relation. If I take
the Hughes slope, however, the average masses rise to $1.1 \times 10^{18}$g and $1.2 \times 10^{18}$g 
respectively. These values translate into total outer Oort cloud masses of 2 -- 40 Earth masses.

\subsection{Terrestrial Impact Probabilities}

My model can be used to calculate the probability of a long-period comet impacting the Earth.
\citet{ste93} calculated that the probability of a given long-period comet with $q \le 1$AU impacting 
the Earth is $\sim 3 \times 10^{-9}$ per perihelion passage. My model suggests that the flux of
comets brighter than $H=19$ with $q<1$AU is 8 per year. The mean time between comet
collisions with the Earth is thus $\sim 40$ million years: very comparable to the figure calculated
by \citet{sek86}, and to the mean time between global extinction events.

Most of these comets will, however, be quite small. Using the conversion between absolute
magnitude and radius described in \S~\ref{small}, a comet with $H \sim 19$ has a radius of
only $\sim 20$m: too small to cause a global extinction event. Comets with radii of
1km or greater ($H \lesssim 9$) are rarer - the mean time between collisions with comets
this large is $\sim 150$ million years.

The shallow faint end slope means that collisions with even small long-period comets
are rare. If, for example, the Tunguska impact was caused by a comet \citep[eg.][]{bro00}, it 
would have
a mass of $\sim 10^{11}$g \citep{vas98} and hence an absolute magnitude of $\sim 18$. The
probability of such a long-period comet impacting the Earth in the last 100 years is thus 
$< 10^{-5}$. The Tunguska impactor must therefore be either asteroidal \citep[eg.][]{sek98,far01} or
associated with a {\em short-period} comet \citep[eg.][]{ash98}.

\subsection{Detection of Interstellar Comets}

No comet has ever been detected with a strongly hyperbolic original orbit \citep{kre92}: ie. 
a comet that was not gravitationally bound to the solar system.
Several authors have discussed this \citep[eg.][]{sek76,mcg89,ste90,sen93}, with many 
claiming that this is surprising.
Oort cloud formation models predict that for every comet that reaches the classical outer Oort
cloud, a factor $\eta = 3$---100 more are expelled into interstellar space \citep[eg.][]{dun87}. 
If most stars have planets, and if planetary formation is usually accompanied by comet ejection, then 
there should
be a substantial population of free-floating interstellar comets. By some estimates, we should
have expected to have seen one or more such comets by now, passing through the inner solar system.

The results in this paper impact upon this question in two ways. Firstly, the non-detection 
of interstellar comets in the
LINEAR sample places an upper limit on their space density. Secondly, 
most previous estimates of the expected number of interstellar comets 
relied upon the \citet{eve67} comet flux: our lower comet flux thus leads to smaller predictions
of the interstellar comet density.

What limit can we place upon the space density of interstellar comets (those with strongly hyperbolic 
orbits) from the non-detection of any by LINEAR? I will assume that interstellar
comets have the same absolute magnitude distribution that I derive for long period comets, and 
that their apparent
brightness varies with heliocentric distance in the same way as dynamically new inbound comets
in our model. Given these assumptions, and an assumed magnitude limit of 19 (point source
equivalent), one can derive the distance $r$ out to which a comet with any given absolute magnitude
could have been detected. To convert this into the volume surveyed during the three years of
the survey, one must allow for the motion of the comets with respect to the solar system, which 
can carry new comets into range. Typical motions of nearby stars with respect to the Sun are 
$v \sim 40 {\rm km\ s}^{-1}$ \citep{gar01}. The volume surveyed in a survey of duration $T$ is thus
\begin{equation}
V = \frac{4}{3}\pi r^3 + \pi r^2 Tv.
\end{equation}
The product of this equation and the absolute magnitude distribution (Equation~\ref{heq}) was
integrated to calculate the number of comets potentially within LINEAR's magnitude limit, for
a given assumed space density of interstellar comets (defined as the number of interstellar 
comets per cubic astronomical unit brighter than $H=11$). The integral suggests that the bulk of
interstellar comets detected will be those with absolute magnitudes near the break at $H = 6.5$.
The results are quite sensitive to the adopted bright-end slope of the absolute magnitude
distribution, being 40\% lower for the Everhart slope as compared to the Hughes slope.

Not all of these comets will be seen: my model suggests that LINEAR finds $\sim 70$\% of the
Oort cloud comets within its magnitude limit. The fraction may be lower for interstellar comets
because they move faster and hence are not observable for long, but the velocity difference
is only $< 50$\%. Furthermore, a larger fraction of interstellar comets will be brght ones
seen at large
heliocentric distances, where the visibility period is larger. I adopt a conservative 50\% detection
probability, which should be ample to include comets not being followed up or not having good
orbit determinations. The mean number of comets seen over the three years is then evaluated as 
a function
of the assumed average density. If more than 5 comets are predicted to have been observed, the
Poisson probability of us having not seen any interstellar comets is less than 5\%: this is
our adopted limit.

I thus derive an upper limit on the local space density of interstellar comets of
$6 \times 10^{-4}$ per cubic AU (95\% confidence) if the Hughes bright-end slope is
assumed. For the Everhart bright-end slope, this limit increases to $9 \times 10^{-4}$ per 
cubic AU. These limits are very comparable to the best existing limit: that of \citet{sek76}.
Sekanina's limit is, however, based upon Everhart's papers and should thus be regarded with
some suspicion. I can extend this calculation by noting that LINEAR has not discovered any 
interstellar comets in other years. From 1999 through to the end of 2004 their monthly sky
coverage (though not available in detail) is at least comparable to that during my sample
period. These extra three years of data reduce our upper limits to 3 --- 4.5 $\times 10^{-4}$
per cubit AU.

I now evaluate the expected space density of interstellar comets, given our reduced
Oort cloud population estimate. Following \citet{ste90}, the number density of interstellar
comets $n_{\rm ism}$ is given by:
\begin{equation}
n_{\rm ism} = \eta \ n_{\rm stars} N_{\rm comets}, 
\end{equation}
where $n_{\rm stars}$ is the local number density of stars, $N_{\rm comets}$ is the 
mean number of outer Oort cloud comets per star, and $\eta$ is the ratio of comets expelled
from a solar system to the number ending up in the outer Oort cloud.

I adopt $n_{\rm stars} \sim 0.1$ per cubic parsec, from the 8pc sample of \citet{rei97}: this is
consistent with the value used by \citet{mcg89} but considerably lower than the value used
by \citet{ste90}.
For $N_{\rm comets}$, I assume that the number of comets generated per star is proportional
to its mass and metallicity. This may or may not be true, but is consistent with the
observed tendency for low mass stars to lack hot Jupiter planets \citep[eg.][and refs therein]{lin03}. 
The local stellar population is dominated by low mass dwarfs: the average
stellar mass of the nearby stars in the \citet{rei97} catalogue is only $\sim 0.3 M_{\Sun}$.
The average metallicity of near-by F and G stars is $[Fe/H] \sim -0.14$ \citep{nor04}:
no information is available for local dwarf stars, so I assume the same value. Assuming
that the outer Oort cloud population derived in \S~\ref{number} is typical of local stars
with solar mass and metallicity, I therefore derive $N_{\rm comets} = 0.3 \times 0.72 \times 
2 \times 10^{11} = 4.3 \times 10^{10}$. The most uncertain parameter is $\eta$: literature
values range from 3 -- 100 \citep[eg.][]{dun87,mcg89,ste90,wei96}.

Given these parameters, $ 1.4 \times 10^{-6}  < n_{\rm ism} < 4.7 \times 10^{-5}$. Thus even
in the most optimistic case, the predicted flux is an order of magnitude 
below current limits.

Could future surveys reach these predicted densities? I ran the simulation for five year surveys
reaching to deeper magnitude limits. A survey reaching 24th magnitude (perhaps PanStarrs) 
would place 95\%
upper limits of $\sim 5 \times 10^{-5}$, and might hence detect one or two interstellar
comets if $\eta$ is very large. \citet{jew03} reached similar conclusions.
A survey reaching 26th magnitude (LSST?) would push the limit
down to $\sim 1.5 \times 10^{-5}$, which would be enough to detect or rule out a large
value of $\eta$. This prediction does, however, depend crucially on the assumed brightness
behavior of comets a long way from the Sun.

\subsection{Perihelion Distribution}

Simple models which assume that Oort cloud comets have an isotropic velocity
dispersion imply a constant number of comets per unit perihelion. $n$-body integrations
\citep[eg.][]{tsu92,wei99}, in contrast, imply a rising number of comets at larger
perihelia, due to the diffusion process of comets past the perturbations of the giant planets.
Both \citet{tsu92} and \citet{wei99}, for example, predict an increase in the number of 
comets per unit perihelion of $\sim \times 2$ between $q=2$ and $q = 8$ (though this is 
an extrapolation of the Wiegert \& Tremaine model which is only shown out to $q=3$). Is 
this consistent with the LINEAR sample?

The best-fit model has the number of comets per unit perihelion (beyond $q = 2$) going as a power
law of index $\gamma = -0.27 \pm 0.3$ (95\% confidence), ie. a gentle {\em fall}. To get a 
rise in numbers consistent
with the \citet{tsu92} and \citet{wei99} predictions, we require $\gamma \sim 0.5$, which is 
inconsistent with our model with 99\% confidence.

The measured perihelion distribution is, however, somewhat degenerate with the assumed
dependence of comet brightness on heliocentric distance: $n$ in equation~\ref{eqabsmag}. 
As I have repeatedly noted, the choice of $n$ is an approximation based on extrapolations of
observations obtained at much smaller heliocentric distances.

We can estimate the change in $n$ that would be needed to bring our observations
into line with the \citet{tsu92} predictions. We need to drop the predicted brightnesses of
comets with $q \sim 8$ by enough to reduce the observed numbers by 50\%, to meet our
95\% upper limit. Given our best-fit absolute magnitude relation, this requires that
$n$ be increased by $\Delta n > 0.8$.

This is quite a small rise - well within the observed scatter of $n$ values seen at lower
perihelia. If, for example, I had used the canonical value of $n=4$ for all comets, 
rather than our more complex scheme, this would make dynamically new comets much
fainter when far from the Sun, as required (at the expense of the correlation seen in 
Fig~\ref{q_a}).
The data in this paper, while suggestive, are not therefore significantly at
odds with the theoretical predictions. A better understanding of heliocentric brightness
variations when distant from the Sun will be needed to see if this anomaly is real.

\subsection{Implications of the Shallow Faint-End Slope\label{small}}

The faint-end slope of the absolute magnitude relation derived in this paper
($1.01 \pm 0.11$) is very flat:
the number of comets per unit magnitude barely increases as you go fainter. 
This presumably
indicates that small comets are not that much more abundant than large ones. 
If the absolute
magnitudes are converted to nuclear masses, using either the \citet{bai88} or \citet{wei96}
relations, I find that the differential number of comets per unit mass $m$ goes as:
\begin{equation}
\frac{dn}{dm} \propto m^{-1.04 \pm 0.1}.
\end{equation}

\citet{bra96} made the case for the existence of small comets: those with nuclei only meters to
tens of meters in radius. If we assume the \citet{wei96} relationship between mass and
absolute magnitude, and a density of $0.6 {\rm g\ cm}^{-3}$, 100m radius corresponds to
$H \sim 15$ and 10m to $H \sim 20$. LINEAR is therefore detecting at least a few
comets with nuclei smaller than 100m. Similar small comets are also detected by the LASCO
instrument on the SOHO spacecraft \citep{bie02}, though these are mostly fragments of recently
disintegrated larger comets \citep{sek04}.

\citet{hug01}
was unclear on whether the shallow faint-end slope was real or a selection effect: I confirm
that it is real. \citet{sek86} pointed out that small long-period comets must be rare from 
statistics of
comets passing close to the Earth. \citet{mee04} found a similar paucity of small Jupiter-family 
comets. 

In this section, I point out that the shallowness of the faint-end slope is actually quite 
interesting theoretically. Collisions are rare in the Oort cloud \citep{ste88}, so the
nuclear size distribution should remain largely unchanged from when the proto-comets were
planetesimals expelled from the protoplanetary disk \citep{dun87,gol04}.

\citet{wei97} modeled the size distribution of planetesimals in the protoplanetary disk.
These models imply that 100m diameter objects should outnumber 2km sized objects by a factor
of $\sim 10^{10}$ (per unit log radius) in the oldest disks modeled. Our faint-end slope,
however, if combined with the \citet{wei96} mass/absolute magnitude relation, implies a ratio
of only $\sim 10^4$.

The size distribution of planetesimals may be greatly modified by collisions while
they lie within the dense environment of the proto-planetary disk. Most models of
this collisional evolution also, however, predict much flatter size distributions than we
see: they typically predict an increase in comet numbers, even at faint magnitudes,
of a factor of $\sim 3.2$ per magnitude \citep{doh69,wet93,cha03}. This is much greater than
my measurement ($1.03 \pm 0.09$).

As noted in \S~\ref{eve}, our shallow faint-end slope is quite robust to sample
incompletenesses and model assumptions. The conversion of absolute magnitudes to radii is,
however, highly uncertain even at bright magnitudes, let alone the faint absolute magnitudes
of relevance here.  Theoretical predictions are also quite uncertain 
\citep{lis93,you02}, and may be consistent with this slope. Another possibility, however,
is that the probability of a planetesimal escaping the protoplanetary disk and reaching
the Oort cloud is size-dependent. \citet{ste01}, for example, suggested that collisions 
between planetesimals would act to circularize their
orbits, and would prevent them from escaping into the Oort cloud until the density of
planetesimals was greatly depleted \citep[see also][]{cha03}. The square-cube law would 
suggest that this effect is most serious for smaller planetesimals, which may  
be ground down to dust before
escaping. This could thus explain the deficit of small comets. Alternatively, gas
drag could play the same role: the largely gaseous nature of the giant planets
indicates that the protoplanetary disk was still full of gas when the giant planets
formed. No published modelling currently includes this effect.

Another possibility: the square-cube law implies that small comets may loose their 
volatiles faster than large ones. Fading may thus be more severe for these comets, making 
them harder to detect.
A final possibility is that comets have been exposed to high temperatures at some
point in their history. The resultant loss of volatiles could thus destroy small comets
without much affecting the numbers of large ones. \citet{mee04} tentatively suggest this as
a reason for the lack of Jupiter-family comets with small nuclei. Long period comets might have 
been exposed to high temperatures while still in the proto-planetary disk. Once in the Oort cloud,
temperatures are much lower. Nearby supernovae and O-stars may have temporarily heated them 
enough to remove some volatiles
\citep{ste03}, but this should not appreciably affect the numbers of small comets.

I note in passing that my models are strongly inconsistent with the claims by \citet{fra01}
for an immense population of small comets bombarding the Earth. These claims have, however, already
been ruled out in many other ways \citep[eg.][]{bas97,han99,kno99,har00}.

\subsection{Implications for Future Surveys}

The model in this paper can be used to guide future automated comet-searches. The most basic 
conclusion follows from the shallow faint-end slope of the number/absolute magnitude relation.
This means that to find more comets, a survey should always try to maximize the area covered 
rather than going deep in a small area. If LINEAR, for example, exposed for six times as long
per field, it would be sensitive to fainter comets, and would hence see $\sim 1.5$ times
more comets per unit area. But it would cover a six times smaller area.

\begin{figure*}
\plotone{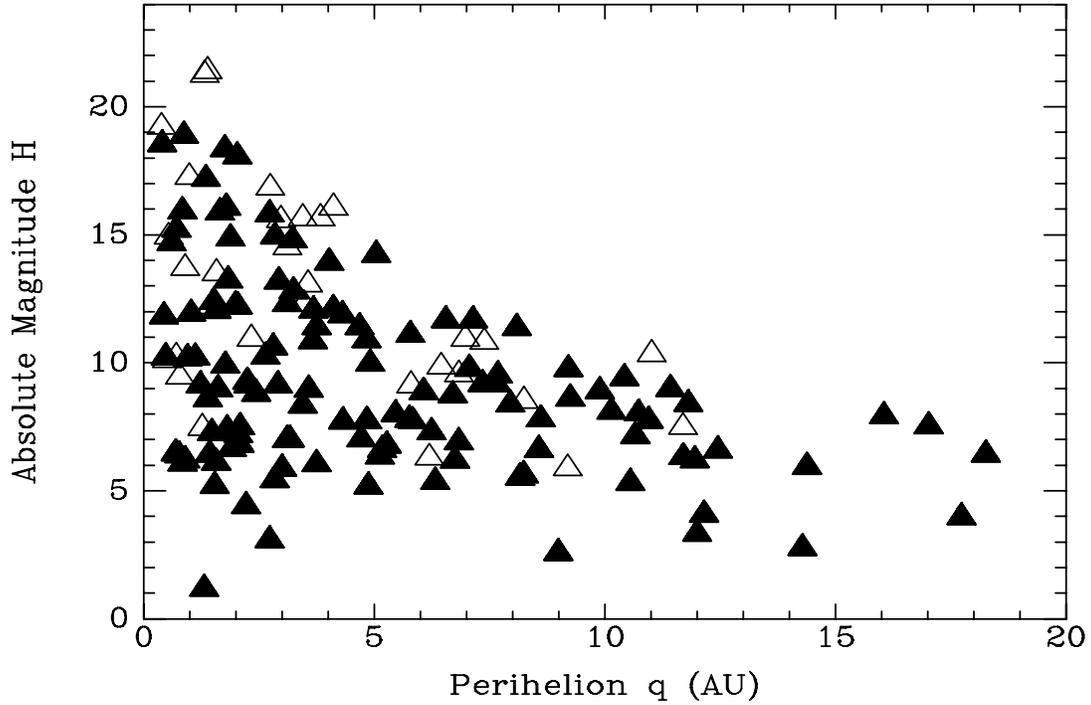}
\caption{
The predicted long-period comet sample found by a hypothetical survey regularly covering the 
whole accessible sky down to a point-source magnitude limit of 22. Open triangles are the
comets that would be found if this region were surveyed once per week. Solid triangles are those
that would also have been seen if the region were only surveyed twice a year. The sample was 
obtained for a hypothetical three year survey and only includes comets reaching perihelion
within these three years. No weather losses were included.
\label{freq}}
\end{figure*}

The more frequently a survey covers a given area of sky, the higher the probability of a given
comet being detected. In Fig~\ref{freq}, however, I show that this is not an enormous effect:
decreasing the survey frequency for a hypothetical survey from weekly to biannual only drops the
number of detected comets by $\sim 20$\%. The comets lost are primarily those with fainter
absolute magnitudes, because their visibility period is small. The small number lost is a direct
consequence of the shallow faint-end slope: there are few small comets to lose.

\begin{figure*}
\plotone{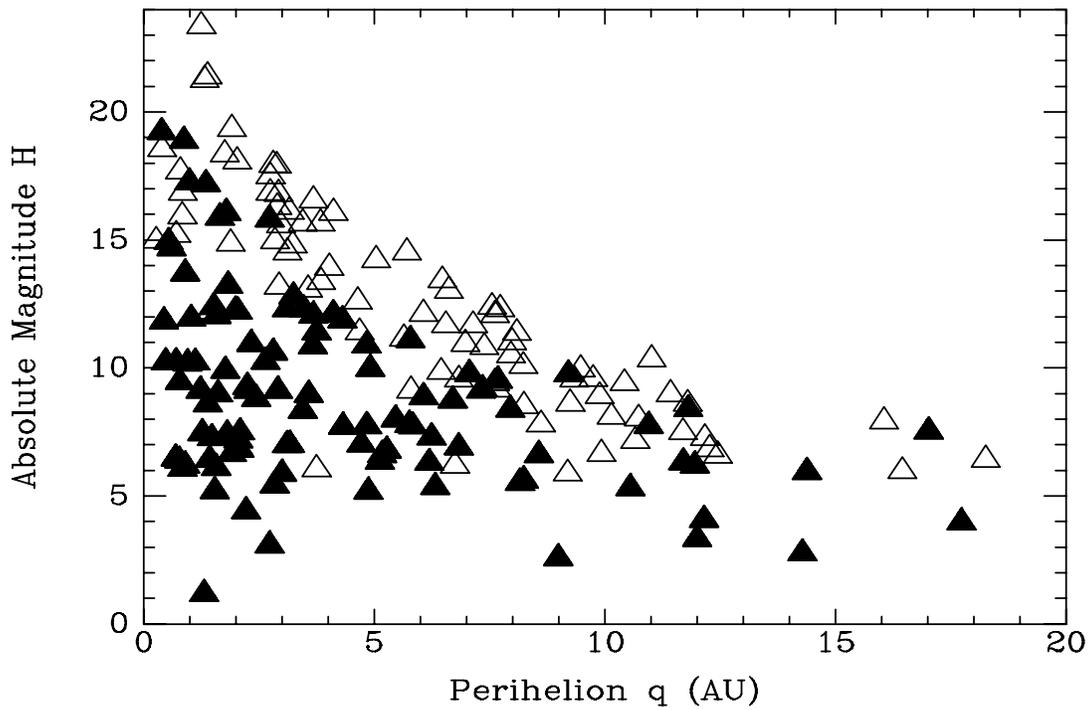}
\caption{
The predicted long-period comet sample found by a hypothetical survey regularly covering the 
whole accessible sky once per month. Open triangles are the comets that would be found if 
survey reached a point-source magnitude limit of 24. Solid triangles are those
that would also have been seen if limit were only 20. The sample was obtained for a 
hypothetical three year survey and only includes comets reaching perihelion
within these three years. No weather losses were included.
\label{faint}}
\end{figure*}

In Fig~\ref{faint}, I show the predicted samples that would be found by telescopes using 
a similar survey technique to LINEAR, but going deeper. The deeper limit essentially
increases the absolute magnitude limit reached at all perihelia, and by increasing 
the time over which comets are visible, also improves the completeness for brighter 
comets. Over 3 years, a survey to 20th mag would detect 103 comets, to 22nd mag would 
detect 150 and to 24th mag, 186 comets. This is much fewer than \citet{jew03} estimated.
The discrepancy is probably due to my shallow inferred absolute magnitude and perihelion
distributions, which mean that the dramatic increase in sensitivity of these surveys as
compared to LINEAR only yields a relatively small increase in sample size.

I conclude that telescopes such as SkyMapper, PanStarrs and LSST, each capable of surveying large 
areas to deeper than 22nd magnitude, should be capable of detecting
more than 50 long-period comets per year. A substantial fraction of these are forecast to
have perihelia beyond 10AU, though this conclusion relies upon the rather shaky assumptions of
how brightness varies with heliocentric distance this far out. Thus five years should suffice to 
build up a quantitatively selected sample of long-period comets equal in size to any historical 
sample.

\section{Conclusions}

\label{conc}

Six main conclusions can be drawn from this analysis:

\begin{enumerate}

\item The outer Oort cloud contains $\sim 5 \times 10^{11}$ comets down to $H=17$ ($\sim 2 \times 
10^{11}$ comets down to $H=11$). This is 2 --- 10 times fewer comets than previous published 
estimates.

\item The average mass of these comets is, however, higher than previous estimates. Down to
$H=11$, the average mass is between $5.6 \times 10^{16}$g and $1.2 \times 10^{18}$g, leading to 
a total mass in the outer Oort cloud of 2 --- 40 Earth masses, comparable to or larger than previous
estimates. The mass of the Oort cloud may be dominated by a few large comets.

\item Small comets do exist, but are rarer than predicted by many models. This may be because
they have difficulty escaping from the protoplanetary disk. The probability of the Earth being hit
by a long-period comet similar in energy to the Tunguska impactor is only one in forty million per
year.

\item I place an upper limit on the space density of interstellar comets of 
$4.5 \times 10^{-4}$ per cubic AU (95\% confidence). This is still an order of magnitude
above our revised prediction for the space density of interstellar comets.

\item The number of long-period comets per unit perihelion seems to decline, or at best
rise slowly beyond 2AU. This does not agree with theoretical predictions. The discrepancy
may be resolved if comets are fainter at large heliocentric distances than I assume.

\item Future survey telescopes should be able to assemble samples of several hundred long-period
comets in a few years of operation.

\end{enumerate}

The major weakness in this analysis is in the photometry: in particular in our limited
understanding of how the brightness of comets varies with heliocentric distance when far
from the Sun, and in how to convert ill-defined total magnitudes into more reproducible
and physically meaningful parameters such as $Af \rho$. The data exist to address these problems,
but are not publicly available.

\acknowledgements

I would like to thank Grant Kennedy, Eriita Jones and Chris Weekes for their work on aspects of this
paper, Stephen Pravdo and Grant Stokes for responding to e-mail questions about their NEO
surveys, Timothy Spahr for providing details of objects posted to the NEO confirmation
page, and Dan Green for provide magnitude measurements from the archives of the International Comet
Quarterly. Paul Weissmann was referee on an earlier version of this paper: his detailed comments
were invaluable in educating the author (an extragalactic astronomer by training) in comet lore.

\end{document}